\documentclass[10pt,twocolumn,journal]{IEEEtran}
\usepackage{latexsym}
\usepackage{float}
\usepackage{amsfonts}
\usepackage{amsbsy}
\usepackage{amssymb}
\usepackage{times}
\usepackage{graphicx}
\usepackage{enumerate}
\usepackage[usenames]{color}
\usepackage[dvips]{pstcol}
\usepackage{epstopdf}
\usepackage{cite}
\usepackage{amssymb}
\usepackage{amsfonts}
\usepackage{graphicx}
\usepackage{epsfig}
\usepackage{psfrag}
\usepackage{xcolor}
\usepackage{amsfonts, bm}
\usepackage{epstopdf}
\usepackage{cite}
\usepackage{color}
\usepackage{xcolor}
\usepackage{subfig}
\usepackage{verbatim}
\usepackage{multirow}
\usepackage{array}
\usepackage{booktabs}
\usepackage{amsthm}

\usepackage[linesnumbered, ruled]{algorithm2e}
\usepackage{algpseudocode}
\usepackage{amsmath}

\newtheorem{remark}{Remark}

\IEEEoverridecommandlockouts
\begin{document}
\title{ Robust Semantic Communications with Masked VQ-VAE Enabled Codebook  }
\author{ Qiyu Hu, \textit{Student Member, IEEE}, Guangyi Zhang, \textit{Student Member, IEEE}, Zhijin Qin, \textit{Senior Member, IEEE}, \\ Yunlong Cai, \textit{Senior Member, IEEE}, Guanding Yu, \textit{Senior Member, IEEE}, and Geoffrey Ye Li, \textit{Fellow, IEEE}
\thanks{ Q. Hu, G. Zhang, Y. Cai, and G. Yu are with the College of Information Science and Electronic Engineering, Zhejiang University, Hangzhou 310027, China (e-mail: qiyhu@zju.edu.cn; zhangguangyi@zju.edu.cn; ylcai@zju.edu.cn; yuguanding@zju.edu.cn). 

Z. Qin is with the Department of Electronic Engineering, Tsinghua University, Beijing 100084, China (e-mail: qinzhijin@tsinghua.edu.cn).

Geoffrey Ye Li is with the Department of Electrical and Electronic Engineering, Imperial College London, London SW7 2AZ, UK (e-mail: geoffrey.li@imperial.ac.uk).

This work was supported in part by the National Natural Science Foundation of China (NSFC Nos 62293484 and 61925105) and in part by Tsinghua University-China Mobile Communications Group Company Ltd. Joint Institute.
This work was supported in part by the Fundamental Research Funds for the Central Universities 226-2022-00195, the Defense Industrial Technology Development Program under Grant JCKY2020210B021.
Zhejiang Provincial Key Laboratory of Information Processing, Communication and Networking (IPCAN), Hangzhou 310027, China.
  } }

\maketitle
\vspace{-3.5em}
\begin{abstract}
Although semantic communications have exhibited satisfactory performance on a large number of tasks, the impact of semantic noise and the robustness of the systems have not been well investigated. Semantic noise refers to the misleading between the intended semantic symbols and received ones, thus causes the failure of tasks. In this paper, we first propose a framework for the robust end-to-end semantic communication systems to combat the semantic noise. In particular, we analyze sample-dependent and sample-independent semantic noise. To combat the semantic noise, the adversarial training with weight perturbation is developed to incorporate the samples with semantic noise in the training dataset. Then, we propose to mask a portion of the input, where the semantic noise appears frequently, and design the masked vector quantized-variational autoencoder (VQ-VAE) with the noise-related masking strategy. We use a discrete codebook shared by the transmitter and the receiver for encoded feature representation. To further improve the system robustness, we develop a feature importance module (FIM) to suppress the noise-related and task-unrelated features. Thus, the transmitter simply needs to transmit the indices of these important task-related features in the codebook.  
Simulation results show that the proposed method can be applied in many downstream tasks and significantly improve the robustness against semantic noise with remarkable reduction on the transmission overhead.  
\end{abstract}


\IEEEpeerreviewmaketitle

\section{Introduction}
Traditional communication systems focus on efficient symbol transmission and accurate symbol recovery \cite{ConvenCom}, where symbol-error rate (SER) and bit-error rate (BER) are usually used as the performance metrics. Various new applications generate unprecedented amounts of data for serving different types of tasks while the conventional communication system is facing the bottleneck to support such massive amount of data \cite{IoTData}. Moreover, it causes critical challenges for conventional communications to support massive connectivity over limited spectrum resources but with low latency \cite{Roadmap6G}.

\subsection{Prior Work}
With the development of deep neural networks (DNNs) \cite{Roadmap6G} and end-to-end learning \cite{E2EGAN,E2ELearning,E2EIntro}, semantic communications, which extract and transmit the task-related meanings of data \cite{SemEmpower,GoalOriented,SemanMaga,JSCC}, have emerged as a key technology and received great attention. For example, in the image transmission for object detection task, the location and shape of an object are task-related while the background is unrelated to the task and unnecessary to transmit. Moreover, semantic communication is robust to unfriendly channel environments, i.e., low signal-to-noise ratio (SNR), which fits well the applications requiring high reliability \cite{DeepSC}. These advantages motivate us to develop communication systems by considering the semantic meaning beyond digital bits to enhance transmission accuracy and efficiency. 

The existing works on semantic communications can be divided into two categories: data reconstruction \cite{JSCC,DynamicSNR,DeepSC,DRLSemantic,JSCCf,Speech,HARQ} and task execution \cite{TransIoT,ImagRetri,IBClass,VQA,UNIT}. For the data reconstruction, the semantic information behind the data is extracted and only the related data is reconstructed based on the received semantic information. The joint semantic-channel coding (JSCC) scheme in \cite{DeepSC} extracts the semantic information from text. In \cite{DRLSemantic}, reinforcement learning is exploited to recover the text. The semantic communication system in \cite{JSCCf} with channel feedback can improve the quality of image reconstruction. The attention-based semantic communication system in \cite{Speech} focuses on speech signals.

For the task-specific applications, the task-related semantic information encoded at the transmitter is directly applied for task execution at the receiver. In particular, the image classification-oriented semantic communications in \cite{TransIoT} can improve the recognition accuracy. The task of image re-identification for a person in \cite{ImagRetri} can enhance the retrieval accuracy. The semantic communication system in \cite{IBClass} considers edge inference with classification task. The semantic communication system in \cite{VQA} is designed for the multi-modal data task: the visual question answering. 


Though the aforementioned deep learning (DL)-based semantic communication systems have exhibited very impressive performance in certain tasks, the impact of noise and the system robustness still need to be further investigated. There have been studies on analyzing the generation and characteristics of different kinds of image and text noise and a number of denoising algorithms have been proposed \cite{ImageNoise}. However, there exists a particular type of noise in semantic communications that has not been well studied \cite{GoalOriented}. 
More or less similar to the noise in the conventional systems, semantic noise causes misunderstanding of semantic information and decoding errors, which results in misleading between the intended semantic meaning and the reconstructed one at the receiver. The semantic noise can be generated in different stages, including semantic encoding, data transmission, and decoding \cite{SemanMaga}. 
In the semantic encoding stage, the semantic noise corresponds to the mismatch between the original signal and the encoded signal after semantic encoding, which is related to the representational capability of the encoder. In the data transmission stage, the signal distortion caused by channel fading and some well-designed signals sent by malicious attackers both introduce semantic noise. 
In the decoding stage, the misinterpretation, incorrect representation, and confusion of meanings introduce semantic noise to the receiver. For example, in the process of semantic symbol recognition, the same semantic symbol is employed to represent different sets of data with different meanings when the reconstructed symbols are ambiguous.

Semantic noises for different categories of sources, e.g., text and image, are generally different \cite{Principle}. The semantic noise in text refers to semantic ambiguity, where slight changes to words in the sentence, e.g., synonym replacement or randomly reverse alphabetical order, may make the DL model misunderstand the semantic meaning of the sentence \cite{texts}. The semantic noise in the image can be modeled based on the adversarial samples \cite{intriguing}, which is different from that in the text. Due to the discrete nature of text, it is impossible to add perturbation to the text without being noticed by humans. However, some subtle modification can be added to the images that is barely noticeable to humans. Adversarial samples mislead DL models and cause significant performance degradation, but they look identical to the original images for humans.  

In this paper, we focus on the semantic noise in the image domain and that in the text domain can be handled in a similar way. The methods for generating adversarial samples in the image domain can be classified into two categories: (i) the sample-dependent method, which fools a DNN on a single image \cite{FGSM,PGD,limitations,deepfool}; (ii) the sample-independent universal method, which fools a DNN on any image with a high probability \cite{Universal,UniversalCSI}. In particular, the methods on the first category include the fast gradient sign method (FGSM) \cite{FGSM}, projected gradient descent (PGD) \cite{PGD}, Jacobian-based saliency map attack (JSMA) \cite{limitations}, and deepfool algorithm \cite{deepfool}. 
The existing DL-enabled systems are vulnerable and particularly unstable to these adversarial samples, where small and imperceptible perturbations of the data samples are sufficient to fool them and would result in incorrect results \cite{intriguing}. To improve the robustness of DL models against adversarial samples, there have been some methods, such as input denoising \cite{defense}, defensive distillation \cite{distillation}, gradient regularization \cite{gradient}, weight perturbation \cite{WeightPerturb}, and adversarial training \cite{advertrain}. 

\subsection{Motivation and Contributions}

It is foreseen that the robustness of semantic communications in terms of security and reliability is important in future applications, such as self-driving vehicles and medical diagnosis. Although there have been a number of methods for generating adversarial perturbations in the field of image processing, the semantic noise model in wireless communications has not been well investigated. 
Moreover, the performance of these methods against adversarial perturbations is unsatisfactory and even deteriorates in the clean samples without noise. More importantly, the impacts of wireless channels and transmission overhead in communications are usually ignored. Therefore, in this paper, we model the semantic noise in the communication field and design a robust semantic communication system that can effectively combat semantic noise with lower transmission overhead. 
We propose a DL-enabled end-to-end robust semantic communication system to combat the semantic noise. Both the transmitter and receiver are represented by DNNs. Since it is a data-driven method without pre-assumed channel models as a prerequisite, it can potentially provide a solution with high generalization capability to various communication scenarios.  

We firstly model the semantic noise in practical wireless communication environments. In particular, we employ the iterative FGSM method to generate the sample-dependent semantic noise at the transmitter instantly, which adds different semantic noise at each image. Since it is difficult to acquire each channel and the transmitted signal, we further propose an iterative method to generate the sample-independent semantic noise at the receiver. It adds the same semantic noise to different transmitted images and fools most images without requiring the channel state information (CSI) and the transmitted images. To combat the semantic noise, we propose an adversarial training method with weight perturbation, which incorporates the samples with semantic noise in the training dataset to solve a complicated min-max optimization problem. Then, the masked vector quantized-variational autoencoder (VQ-VAE) with vision Transformer (ViT) blocks \cite{Transformer,MAE} is designed as the architecture of the robust semantic communication system. A novel strategy is proposed to mask a portion of the original image, where the semantic noise appears with a high probability. Moreover, a discrete codebook shared by the transmitter and the receiver is designed for encoded feature representation. It focuses on the task-related feature representation and neglects imperceptible noise-related details, which reduces the impact of semantic noise. 

To further improve the system robustness, we design a feature importance module (FIM) that dynamically learns and incorporates the feature importance to the masked VQ-VAE. It inherently suppresses the task-unrelated and noise-related features. Thus, the transmitter only needs to send the indices of the important task-related features in the codebook. Furthermore, the SNR is incorporated into the FIM, which ensures that the proposed system can successfully operate in a wide range of SNR levels. 
Moreover, existing works in semantic communications focus on mapping the source data directly into channel symbols for transmission. It is called full-resolution constellation because the constellation points can appear anywhere in the constellation. However, this is difficult to realize for practical systems due to the finite precision and might be impractical in the current digital communication systems \cite{JSCC,DeepSC,DynamicSNR}. Our proposed masked VQ-VAE model with FIM can design a discrete codebook-based system, which is more practical and can be easily realized in the current digital communication systems since the indices of the features can be directly mapped into symbols via employing the existing constellation. 
Simulation results show that our proposed method can be applied in many downstream tasks and can significantly improve the robustness of semantic communication systems against semantic noise with much reduced transmission overhead. The main contributions of this paper are summarized as follows:
\begin{itemize}
\item Based on adversarial perturbations in computer vision \cite{Universal}, taking into account the modulation, channel, and demodulation in communication systems, we model the sample-dependent and sample-independent semantic noise added at the transmitter and receiver, respectively.

\item Based on the basic adversarial training method \cite{PGD} and optimization theory, we propose an adversarial training method with weight perturbation to combat the semantic noise, with the consideration of the effects caused by channel impairments in wireless communication systems. 

\item We develop a masked VQ-VAE model with a masking strategy as the architecture of the robust semantic communication system. A discrete codebook shared by the transmitter and receiver is designed for encoded feature representation, which fits the current digital communication system well.

\item We provide performance analysis and propose a novel loss function to improve system robustness based on semantic similarity.

\item To further improve the system robustness, we design the FIM to suppress the noise-related and task-unrelated features.  
\end{itemize}

\subsection{Organization and Notations}
The rest of paper is structured as follows. Section \ref{System} models the semantic noise and proposes a general framework of the semantic communication system to combat the semantic noise. Section \ref{Architect} designs the masked VQ-VAE with a masking strategy and a discrete codebook for encoded feature representation. Section \ref{Efficient} improves the semantic communication systems for stronger robustness by designing the FIM with dynamic SNR. The simulation results are presented in Section \ref{Simulation}. Finally, the paper is concluded in Section \ref{Conclusion}.

\emph{Notations:} Scalars, vectors, and matrices are respectively denoted by lower case, boldface lower case, and boldface upper case letters. Notation $\mathbf{I}$ represents an identity matrix and $\mathbf{0}$ denotes an all-zero matrix. For a matrix $\mathbf{A}$, ${\bf{A}}^T$, $\mathbf{A}^*$, ${\bf{A}}^H$, ${\bf{A}}^{-1}$, ${\bf{A}}^{\dagger}$, and $\|\mathbf{A}\|$ are its transpose, conjugate, conjugate transpose, inversion, pseudo-inversion, and Frobenius norm, respectively. For a vector $\mathbf{a}$, $\|\mathbf{a}\|$ is its Euclidean norm. 
Finally, ${\mathbb{C}^{m \times n}}\;({\mathbb{R}^{m \times n}})$ are the space of ${m \times n}$ complex (real) matrices.

\section{Framework of Robust Semantic Communications} \label{System}
In this section, we model the semantic noise and propose the framework of robust end-to-end semantic communication systems with adversarial training to combat the semantic noise.

\subsection{Semantic Communication Systems}

\begin{figure*}[t]
\begin{centering}
\includegraphics[width=0.75\textwidth]{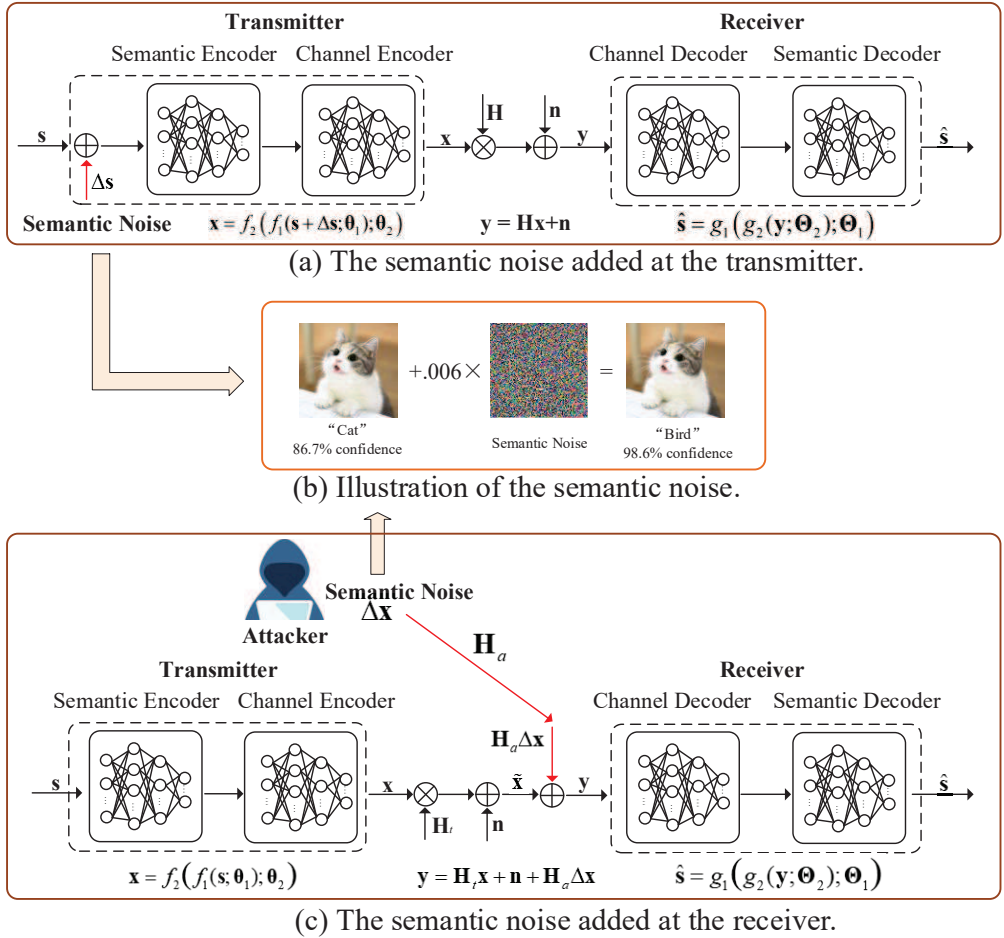}
\par\end{centering}
\caption{The framework of the robust semantic communication system with semantic noise.}
\label{FrameArchitect}
\end{figure*}
  
As shown in Fig. \ref{FrameArchitect}, the transmitter maps the source, $\mathbf{s}$, into a symbol stream, $\mathbf{x}$, and then passes it through the physical channel with transmission impairments. The received symbol stream, $\mathbf{y}$, is decoded at the receiver to obtain an estimation of the source, $\hat{\mathbf{s}}$. Both the transmitter and receiver are represented by DNNs, which are jointly designed. 
In particular, the DNNs at the transmitter consist of the semantic encoder and the channel encoder while the DNNs at the receiver consist of the channel decoder and the semantic decoder. The semantic encoder learns to extract the semantic information from the transmitted data and transforms it into an encoded feature vector while the semantic decoder learns to recover the transmitted data from the received signals. Moreover, the channel encoder and channel decoder aim at eliminating the signal distortion caused by wireless channels.

Assuming that the input is an image, we consider a system with $N_{t}$ transmit antennas and $N_{r}$ receive antennas. The encoded symbol stream can be represented by
\begin{equation} \label{SemCom01}
\mathbf{x}=f_{2}\big( f_{1}(\mathbf{s};\bm{\theta}_{1} ); \bm{\theta}_{2}  \big),
\end{equation}
where $\mathbf{x}\in \mathbb{C}^{N_{t}\times 1}$, $\bm{\theta}_{1}$ and $\bm{\theta}_{2}$ denote the trainable parameters of the semantic encoder, $f_{1}(\cdot)$, and the channel encoder, $f_{2}(\cdot)$, respectively. Subsequently, the received signal, $\mathbf{y}\in \mathbb{C}^{N_{r}\times 1}$, is given by
\begin{equation} \label{SemChannel}
\mathbf{y}=\mathbf{H}\mathbf{x}+\mathbf{n},
\end{equation} 
where $\mathbf{H}\in \mathbb{C}^{N_{r}\times N_{t}}$ denotes the channel matrix and $\mathbf{n} \in \mathbb{C}^{N_{r}\times 1} \sim \mathcal{CN}(\bm{0}, \sigma^{2}\mathbf{I})$ is the additive white Gaussian noise (AWGN). Correspondingly, the decoded signal
is given as
\begin{equation} \label{SemCom02}
\hat{\mathbf{s}}=g_{1}\big( g_{2}(\mathbf{y};\bm{\Theta}_{2} ); \bm{\Theta}_{1}  \big),
\end{equation}
where $\bm{\Theta}_{1}$ and $\bm{\Theta}_{2}$ denote the trainable parameters of the semantic decoder, $g_{1}(\cdot)$, and the channel decoder, $g_{2}(\cdot)$, respectively. For clarity, we denote $\bm{\theta}$ as the trainable parameters and $f_{\bm{\theta}}(\cdot)$ as the DNNs in the considered semantic communication systems. 
Thus, we have $\hat{\mathbf{s}}=f_{\bm{\theta}}(\mathbf{s})$. The goal of this system is to minimize the semantic error while reducing the number of symbols to be transmitted.

\subsection{Generation of Semantic Noise}  
We consider the semantic noise generated at the transmitter and the receiver.

\subsubsection{Semantic Noise at Transmitter} \label{SemanticTX}

This kind of semantic noise is generated in the encoding stage. Consider the scenario where a malicious attacker downloads the image dataset, adds semantic noise to each image, and then uploads the modified dataset. The semantic noise has serious impact on the encoding process and will mislead the DL models to generate wrong results for tasks. However, since the semantic noise is so subtle that the legitimate users barely notice it, they will use these contaminated images as usual. Semantic noise also exists in the nature. For example, images obtained by taking pictures of the adversarial samples could also cause misclassification \cite{Physical}. 

The goal of the semantic communication system is to minimize the loss function for serving a specific task, e.g., the mean square error for image reconstruction and the cross entropy for classification task, etc. In contrast, the semantic noise aims to maximize the loss function. Let $\mathcal{S} = \{ \mathbf{s}_{1}, \mathbf{s}_{2}, \cdots, \mathbf{s}_{I} \}$ be a set of images sampled from the training dataset. Then, the generation of semantic noise for the $i$-th image, $\mathbf{s}_{i}$, can be modeled as the solution of the following optimization problem, 
\begin{subequations}  \label{NoiseProblem}
\begin{eqnarray}
\textrm{P1}: & \max\limits_{\Delta \mathbf{s}_{i}} & \mathcal{L}(f_{\bm{\theta}} (\mathbf{s}_{i} + \Delta \mathbf{s}_{i}), \mathbf{z}_{i}) \\
& \textrm{s.t.} & \|\Delta \mathbf{s}_{i} \|_{p} \leq \epsilon, \label{PowerCons}
\end{eqnarray}
\end{subequations}
where $\mathbf{s}_{i}$ and $f_{\bm{\theta}} (\mathbf{s}_{i} + \Delta \mathbf{s}_{i})$ denote the $i$-th input image and the output of the DNN, respectively, $\Delta \mathbf{s}_{i}$ is the semantic noise generated for the $i$-th image, and $\mathcal{L}(\cdot)$ denotes the loss function of DNN for a specific task. In addition, $\mathbf{z}_{i}$ denotes the target associated with $\mathbf{s}_{i}$, e.g., the true label
for classification task and the original image for image reconstruction task, etc. Note that $\|\cdot \|_{p}$ is the $p$-norm and constraint \eqref{PowerCons} limits the power of semantic noise to avoid being observed by humans. Unless otherwise stated, we select $p=\infty $, i.e., infinite norm, in this paper.

To solve this problem, we employ the FGSM in \cite{FGSM}, which linearizes the loss function as 
\begin{equation}
\! \mathcal{L}(f_{\bm{\theta}} (\mathbf{s}_{i} + \Delta \mathbf{s}_{i}), \mathbf{z}_{i}) \! \approx \! \mathcal{L}( \! f_{\bm{\theta}} ( \mathbf{s}_{i} ), \mathbf{z}_{i}) + (\! \Delta \mathbf{s}_{i} \! )^{T} \nabla_{\mathbf{s}_{i}} \! \mathcal{L}(  f_{\bm{\theta}} ( \mathbf{s}_{i} ), \mathbf{z}_{i} ).
\end{equation}
It is minimized by setting $\Delta \mathbf{s}_{i} = -\alpha \nabla_{\mathbf{s}_{i}} \mathcal{L}( f_{\bm{\theta}} (\mathbf{s}_{i}), \mathbf{z}_{i})$,
where $\alpha$ is a scaling factor to constrain the power of semantic noise to $\epsilon$ in \eqref{PowerCons}. Then, we obtain the semantic noise with power $\epsilon$ as 
\begin{equation}
\Delta \mathbf{s}_{i} = \epsilon \textrm{sign}\big( \nabla_{\mathbf{s}_{i}} \mathcal{L}( f_{\bm{\theta}} (\mathbf{s}_{i}), \mathbf{z}_{i}) \big),
\end{equation}
where $\textrm{sign}(x)=1$ for $x\geq 0$ and $\textrm{sign}(x)=-1$ for $x<0$. Then, the contaminated sample with semantic noise becomes $\mathbf{s}'_{i}=\mathbf{s}_{i}+\Delta \mathbf{s}_{i}$. 
This semantic noise is only generated with one-step iteration of the gradient descent method. To increase the impact of $\Delta \mathbf{s}_{i}$ on the system, we propose to employ the iterative process 
\begin{equation} \label{IteraFGSM}
\mathbf{s}_{i}'^{(k+1)} = \Pi_{\epsilon} \big( \mathbf{s}_{i}'^{(k)} + \alpha \cdot \textrm{sign}( \nabla_{\mathbf{s}_{i}} \mathcal{L}( f_{\bm{\theta}} (\mathbf{s}_{i}'^{(k)}), \mathbf{z}_{i}) ) \big),
\end{equation}
where $k$ denotes the iteration index and $\Pi$ is the projection operator. $\alpha$ is selected to satisfy $K \alpha> \epsilon$ to ensure the full advantage of the noise power $\epsilon$, where $K$ denotes the number of iterations. 

\subsubsection{Semantic Noise at the Receiver}

\begin{algorithm}[t] 
\begin{small}
\caption{Generation of the sample-independent semantic noise at the receiver} 
\label{SemanticRx}
\DontPrintSemicolon
\SetKwInOut{Input}{Input}
\SetKwInOut{Output}{Output}
\SetKwInOut{Initialize}{Initialize}
\Input{The generated channel samples $\{\mathbf{H}_{a}^{(1)}, \mathbf{H}_{a}^{(2)}, \cdots, \mathbf{H}_{a}^{(N)}\}$ and a set of collected received signals $\{\tilde{\mathbf{x}}^{(1)}, \tilde{\mathbf{x}}^{(2)}, \cdots, \tilde{\mathbf{x}}^{(N)}\}$ with their true labels $\{\mathbf{z}^{(1)}, \mathbf{z}^{(2)}, \cdots, \mathbf{z}^{(N)}\}$. The number of iterations $K$, the power constraint $\epsilon$, the scaling factor $\alpha$, the weighting coefficient $\rho$, and the model of decoder. }
\Output{The sample-independent semantic noise $\Delta \mathbf{x}$.}
\Initialize{$\bm{\delta}_{1}=\bm{0}$, $\Delta \mathbf{x}^{(n)}=\bm{0}$, and $\bm{\delta}_{norm}=\bm{0}$. }
\For{$n\leftarrow 1$ \KwTo $N$}{
\% Generate the sample-dependent semantic noise for the $n$-th sample.  \\
\For{$k\leftarrow 1$ \KwTo $K$}{
$\bm{\delta}_{norm}=  \dfrac{ \mathbf{H}_{a}^{(n)*} \nabla_{\tilde{\mathbf{x}}_{k}^{(n)}} \mathcal{L}( g_{d_{\bm{\theta}}} (\tilde{\mathbf{x}}_{k}^{(n)}), \mathbf{z}^{(n)}) }{ \| \mathbf{H}_{a}^{(n)*} \nabla_{\tilde{\mathbf{x}}_{k}^{(n)}} \mathcal{L}( g_{d_{\bm{\theta}}} (\tilde{\mathbf{x}}_{k}^{(n)}), \mathbf{z}^{(n)}) \|_{2} } $; \\ 
$\tilde{\mathbf{x}}_{k+1}^{(n)}=\tilde{\mathbf{x}}_{k}^{(n)}+\alpha \mathbf{H}_{a}^{(n)}\bm{\delta}_{norm}$; \\  $\bm{\delta}_{k+1}=\bm{\delta}_{k}+\alpha \bm{\delta}_{norm}$; \\
}
Weighted average: $\Delta \mathbf{x}^{(n)} = \Delta \mathbf{x}^{(n-1)} + \rho \dfrac{\bm{\delta}_{K}}{\|\bm{\delta}_{K}\|_{2}}$; \\
Normalization: $\Delta \mathbf{x}^{(n)} =\epsilon \dfrac{\Delta \mathbf{x}^{(n)}}{\| \Delta \mathbf{x}^{(n)} \|_{2}}$; \\
}
Output the sample-independent semantic noise $\Delta \mathbf{x}=\Delta \mathbf{x}^{(N)}$.
\end{small}
\end{algorithm}

It corresponds to the semantic noise generated in the transmission and decoding stages, which leads to the failure of decoding and the misunderstanding caused by the interpretation of the receivers. It may come from the non-ideal characteristics of hardware, the signal distortion caused by channel fading, or malicious attackers \cite{SemanMaga,UniversalCSI}. Consider a legitimate transmitter that transmits the encoded signal, $\mathbf{x}$, to the receiver and a malicious attacker that sends the semantic noise, $\Delta \mathbf{x}$, for attacking. Then, the received signal at the receiver is given by
\begin{equation} \label{NoiseRX}
\mathbf{y} = \mathbf{H}_{t} \mathbf{x} + \mathbf{n} + \mathbf{H}_{a} \Delta \mathbf{x} \triangleq \tilde{\mathbf{x}} + \mathbf{H}_{a} \Delta \mathbf{x},
\end{equation}
where $\tilde{\mathbf{x}}\triangleq \mathbf{H}_{t} \mathbf{x} + \mathbf{n}$ denotes the received signal without semantic noise, $\mathbf{H}_{t}$ is the channel between the legitimate transmitter and receiver, $\mathbf{H}_{a}$ denotes the channel between the attacker and receiver, and $\mathbf{n}$ is the AWGN. Note that the semantic noise model in \eqref{NoiseRX} at the receiver is a general model, where $\mathbf{H}_{a}$ can be removed when studying the effects of non-ideal characteristics of hardware and the signal distortion caused by channel fading.

To generate the sample-dependent semantic noise as in Section \ref{SemanticTX}, we need to assume that the attacker knows: (i) the exact channel between the attacker and the receiver, $\mathbf{H}_{a}$; and (ii) the received signal at the receiver, $\tilde{\mathbf{x}}$, in advance, which are not always practical in real wireless communication systems. Therefore, we aim to find a sample-independent semantic noise, $\Delta \mathbf{x}$, that fools most images in the dataset, $\mathcal{S}$, by assuming that the attacker only knows the channel statistics rather than the exact channel. 
In particular, we first generate $N$ channel realizations $\{\mathbf{H}_{a}^{(1)}, \mathbf{H}_{a}^{(2)}, \cdots, \mathbf{H}_{a}^{(N)}\}$ based on the channel statistics and collect a set of received signals $\{\tilde{\mathbf{x}}^{(1)}, \tilde{\mathbf{x}}^{(2)}, \cdots, \tilde{\mathbf{x}}^{(N)}\}$. Then, we generate the sample-independent semantic noise, $\Delta \mathbf{x}^{(n)}$, by using $\tilde{\mathbf{x}}^{(n)}$ and $\mathbf{H}_{a}^{(n)}$ for $n=1, 2, \cdots, N,$ instead of using the real channel. Specifically, we select the scaling factor, $\alpha$, satisfying $K \alpha> \epsilon$ to ensure that we can take full advantage of the noise power, $\epsilon$. To maximize the received power of semantic noise and effectively fool the decoder, the attacker has to fully utilize channel $\mathbf{H}_{a}$. Thus, if semantic noise $\Delta \mathbf{x}$ is multiplied by the conjugate of the channel, $\mathbf{H}_{a}^{*}$, then the received power of semantic noise after going through the channel is maximized.
The generated semantic noise vectors for all the samples, $\{\Delta \mathbf{x}^{(n)} \}$, are weighted averaged and normalized. The details are presented in Algorithm \ref{SemanticRx}, where $\bm{\theta}_{d}$ denotes the parameters of the decoder $g_{\bm{\theta}_{d}}(\cdot)$ and $g_{\bm{\theta}_{d}} (\tilde{\mathbf{x}}^{(n)})$ is the output of the decoder for the $n$-th sample.
In the following, we design the robust semantic communication systems based on the sample-dependent semantic noise model proposed in Section \ref{SemanticTX} and the sample-independent semantic noise can be handled in a similar way.   

\begin{remark}
There exists semantic noise in the original images naturally. Here we take the image classification task as an example, where the ``misunderstanding" caused by semantic noise refers to the ``misclassification", e.g., the transmitter sends an image with a dog but the receiver classifies it as a cat. The misclassification rate would never become zero, due to the inevitable semantic noise that exists in the original images naturally. Moreover, the semantic noise of different datasets is generally different. In particular, the semantic noise power of some simple datasets is low, e.g., MNIST, which is easy to achieve high classification accuracy, while many complicated datasets are with high-power semantic noise, e.g., ImageNet. Besides, the capabilities of deep learning models in handling semantic noise are different. The powerful model with complicated structures can effectively eliminate the semantic noise to achieve better classification accuracy, e.g., ResNet-101. Our proposed semantic noise model strengthens such kind of misunderstanding and requires higher robustness for system.
\end{remark}

\subsection{Adversarial Training}
\subsubsection{Basic Adversarial Training}
The key idea of adversarial training against semantic noise is to add the samples corrupted by the semantic noise into the training dataset \cite{FGSM}.
In particular, the trainable parameters, $\bm{\theta}$, and semantic noise, $\Delta \mathbf{s}_{i}$, are updated iteratively to improve the model robustness. It can be formulated as solving the following min-max optimization problem,
\begin{subequations}  \label{AdvTrain}
\begin{eqnarray}
\textrm{P2}: & \min\limits_{\bm{\theta}} & \frac{1}{I}\sum\limits_{i=1}^{I} \max_{\Delta \mathbf{s}_{i}} \mathcal{L}( f_{\bm{\theta}} (\mathbf{s}_{i}+\Delta \mathbf{s}_{i}), \mathbf{z}_{i} ) \\
& \textrm{s.t.} & \|\Delta \mathbf{s}_{i}\|_{p}\leq \epsilon,
\end{eqnarray}
\end{subequations}
where $I$ denotes the number of training samples. 

To solve (P2), the following two steps are executed iteratively: (i) compute $\mathbf{s}'_{i}$ based on \eqref{IteraFGSM} or Algorithm \ref{SemanticRx} and obtain the semantic noise by $\Delta \mathbf{s}_{i}= \mathbf{s}'_{i}-\mathbf{s}_{i}$. Note that $\bm{\theta}$ is fixed in this step and we add the samples, $\mathbf{s}'_{i}$, into the training dataset; (ii) update $\bm{\theta}$ by the SGD based on the training samples, $\mathbf{s}'_{i}$, to minimize the loss function.

\subsubsection{Adversarial Training with Weight Perturbation}
To further improve the robustness against the semantic noise, we add the weight perturbation, $\bm{\nu}$, on the trainable parameters and reformulate the problem as 
\begin{subequations}  \label{ATAWP}
\begin{eqnarray}
\!\!\!\!\! \textrm{P3}: &\min\limits_{\bm{\theta}} \max\limits_{\bm{\nu}} & \frac{1}{I} \sum\limits_{i=1}^{I} \max\limits_{\Delta \mathbf{s}_{i}} \mathcal{L}( f_{\bm{\theta}+\bm{\nu}} (\mathbf{s}_{i}+\Delta \mathbf{s}_{i}), \mathbf{z}_{i} ) \\
& \textrm{s.t.} & \|\Delta \mathbf{s}_{i}\|_{p}\leq \epsilon, \quad \|\bm{\nu} \|_{p}\leq \gamma \|\bm{\theta} \|_{p}. 
\end{eqnarray}
\end{subequations}
Intuitively, the semantic noise, $\Delta \mathbf{s}_{i}$, and weight perturbation, $\bm{\nu}$, lead to the increase of loss function $\mathcal{L}(\cdot)$ for the $i$-th sample and all the samples, respectively. Thus, the two ``max" operations make solving the inner maximization problem effectively, which results in a better solution of the whole min-max problem \cite{WeightPerturb}. We solve (P3) by Algorithm \ref{minmax}, where the update of weight perturbation $\bm{\nu}$ is
\begin{equation} \label{weightpertur}
\bm{\nu} \leftarrow \Pi_{\gamma} \! \left( \bm{\nu} + \eta \frac{\nabla_{\bm{\nu}} \frac{1}{I} \sum_{i=1}^I \mathcal{L}( f_{\bm{\theta}+\bm{\nu}} ( \mathbf{s}_{i}+\Delta \mathbf{s}_{i} ), \mathbf{z}_{i} ) } {\left\|\nabla_{\bm{\nu}} \frac{1}{I} \sum_{i=1}^I \mathcal{L}(  f_{\bm{\theta}+\bm{\nu}} ( \mathbf{s}_{i}+\Delta \mathbf{s}_{i} ), \mathbf{z}_{i} ) \right\|}\|\bm{\theta}\|  \right),
\end{equation}
which can be derived in a similar way to \eqref{IteraFGSM}. 

\begin{algorithm}[t] 
\begin{small}
\caption{Proposed adversarial training algorithm for solving P3} 
\label{minmax}
\DontPrintSemicolon
\SetKwInOut{Input}{Input}
\SetKwInOut{Output}{Output}
\SetKwInOut{Initialize}{Initialize}
\Input{The training dataset that consists of the input images, $\mathbf{s}_{i}$, with their true labels, $\mathbf{z}_{i}$. }
\Output{The parameters of the trained model, $\bm{\theta}$.}
\For{$m\leftarrow 1$ \KwTo $M$}{
Compute $\mathbf{s}'_{i}$ based on \eqref{IteraFGSM} or Algorithm \ref{SemanticRx} and obtain the semantic noise by $\Delta \mathbf{s}_{i}= \mathbf{s}'_{i}-\mathbf{s}_{i}$. Note that $\bm{\theta}$ and $\bm{\nu}$ are fixed in this step and the computed samples, $\mathbf{s}'_{i}$, are added into the training dataset. \\
Update $\bm{\nu}$ to maximize $\mathcal{L}(\cdot)$ by one step forward and backward propagation in \eqref{weightpertur} with fixed $\bm{\theta}$ and $\Delta \mathbf{s}_{i}$.  \\
Update $\bm{\theta}$ to minimize $\mathcal{L}(\cdot)$ by the SGD with fixed $\bm{\nu}$ based on the training samples $\mathbf{s}'_{i}$. \\  
}
\end{small}
\end{algorithm}

\section{Masked VQ-VAE Enabled Discrete Codebook}  \label{Architect}
In this section, we design the robust semantic communication systems with masked VQ-VAE. A novel masking strategy and a discrete codebook are designed to combat semantic noise with reduced transmission overhead. We provide some performance analysis and propose a novel loss function to improve system robustness based on semantic similarity. 
The proposed codebook is different from that of channel feedback \cite{ConvenCom} in two aspects: (i) we propose a novel masked VQ-VAE to train the codebook together with the encoder and decoder at the transceiver while that of channel feedback is designed by conventional algorithms, e.g., dictionary learning; (ii) we design the codebook for source compression to combat the semantic noise in semantic communications while that of channel feedback is for channel compression in conventional communication systems.

\subsection{Masked VQ-VAE} 
There exists information redundancy in various kinds of sources. The image has heavy spatial redundancy and a missing patch in the image can be recovered from its neighboring patches with the understanding of parts, objects, and scenes. Thus, the strategy of randomly masking partial patches is an efficient approach to create a challenging task that requires the model to build a comprehensive understanding of image statistics and semantic information, which also reduces the information redundancy. Moreover, since the semantic noise is added in the patches of the image, the masking operation can eliminate the effects of semantic noise to some extent.

\subsubsection{The Architecture of Masked VQ-VAE}

\begin{figure*}[!t]
\centering
\subfloat[]{\centering \scalebox{0.30}{\includegraphics{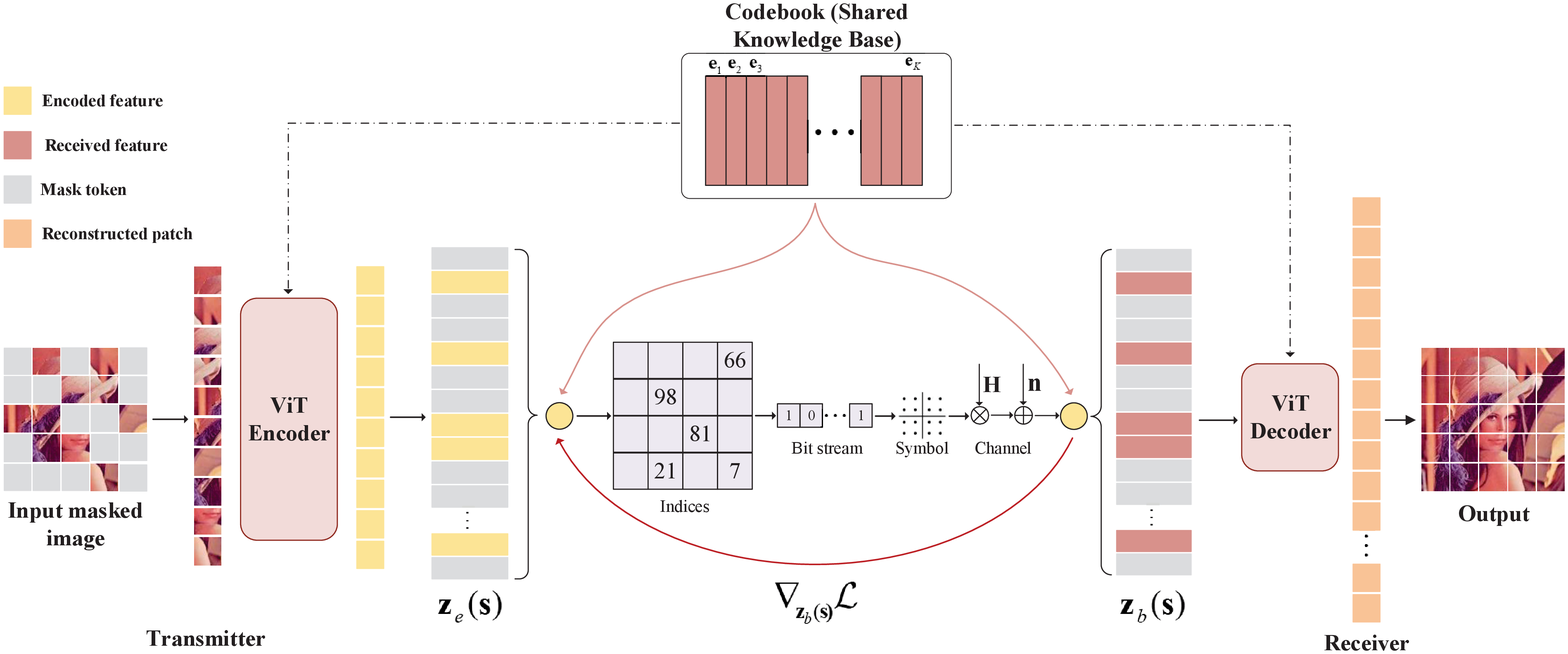}} }
\subfloat[]{\centering \scalebox{0.36}{\includegraphics{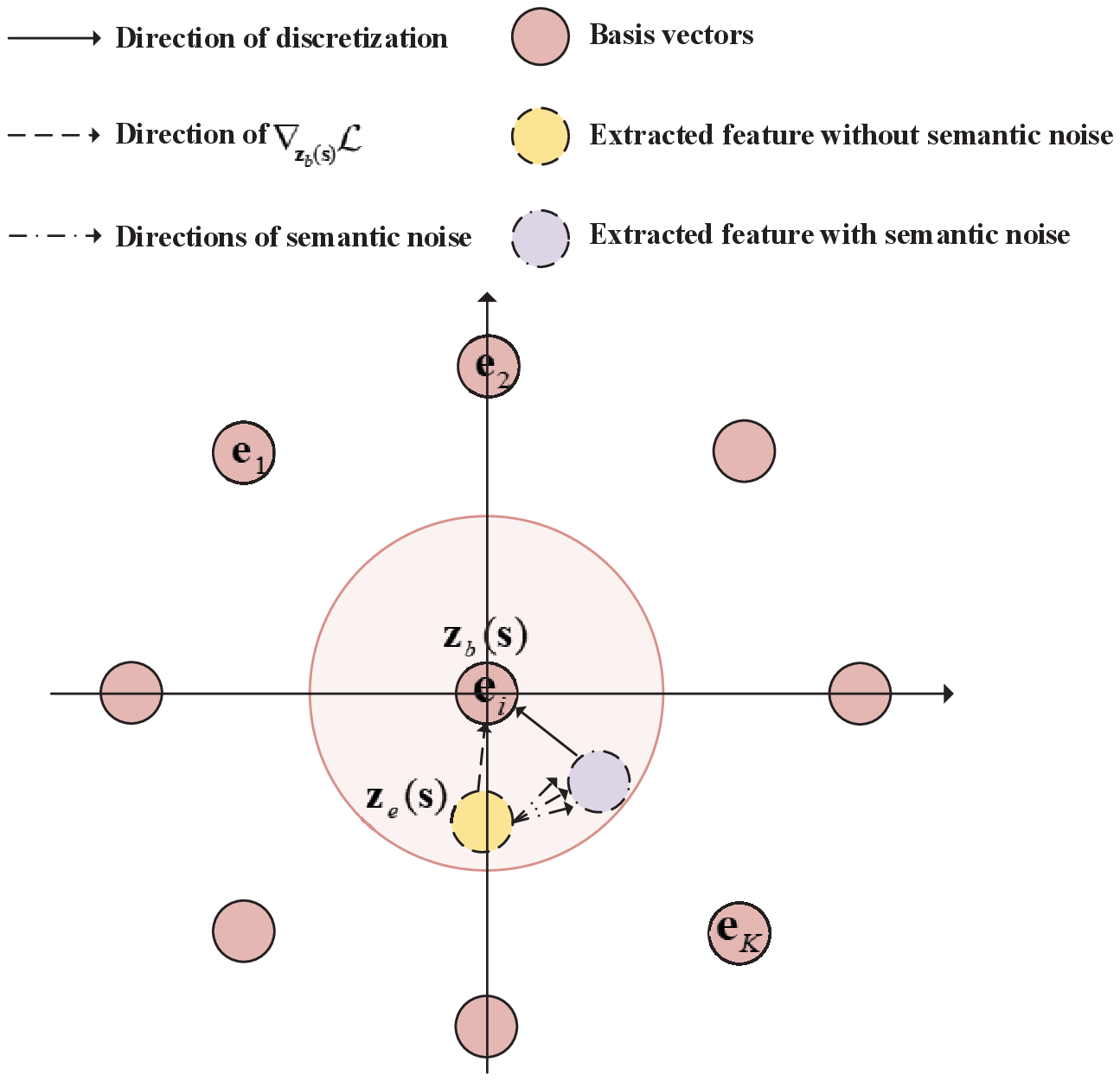}}}
\caption{(a) The architecture of the masked VQ-VAE with discrete codebook; (b) The discrete codebook against semantic noise. }
\label{MAEStruc}
\end{figure*}

As in Fig. \ref{MAEStruc}(a), we employ the masked VQ-VAE with the ViT structure, where we randomly mask patches from the input images and aim to reconstruct the missing patches. The masked VQ-VAE belongs to the autoencoder, but can reconstruct the original image from partial observations. Unlike conventional autoencoders, we adopt an asymmetric encoder-decoder architecture.

In particular, the encoder only needs to process a small portion of the unmasked patches and maps them to the encoded features for transmission, which significantly reduces the training time and memory consumption. It removes the masked patches and embeds the unmasked patches with their positional information in the original image, and then processes them via a series of ViT blocks \cite{Transformer}. 
In contrast, the input to the decoder is the full set of tokens consisting of (i) the encoded features of the unmasked patches and (ii) the mask tokens, as shown in Fig. \ref{MAEStruc}(a). Each mask token is a shared and learned vector that indicates the presence of a missing patch to be predicted. We add positional embeddings to all tokens in this full set. Without this, mask tokens would have no information about their location in the image. Moreover, the decoder is only used during pre-training to perform the image reconstruction task while the encoder is employed to extract the features of the input images. Thus, the decoder architecture, which is independent of the encoder design, can be flexibly designed. Compared with the conventional autoencoder in communication systems, the masked VQ-VAE has the following advantages:
\begin{itemize} 
\item With such an asymmetrical design, the encoder only processes the unmasked patches and the lightweight decoder reconstructs the image from the encoded features and the mask tokens. In this way, the computational complexity and training time can be significantly reduced. 

\item The pre-trained masked VQ-VAE can be employed for different downstream tasks, e.g., classification, simply by changing the structure of the lightweight decoder and fine-tuning the masked VQ-VAE within a short time. 

\item Transmitting the encoded features of the unmasked patches and the mask tokens to the decoder at the receiver leads to a large reduction in transmission overhead. 

\item The masking operation can combat the semantic noise since part of the noise is masked.
\end{itemize}

\subsubsection{Masking Strategy Against Semantic Noise} 
We first divide an image into a number of non-overlapping patches. Then, we sample a subset of patches, mask and remove the remaining patches. A high masking ratio, i.e., the ratio of removed patches, largely eliminates redundancy. The ``random sampling" strategy in \cite{MAE} randomly samples patches following a uniform distribution, i.e., the masking probability of each patch is the same. However, the semantic noise does not appear randomly. It appears in the patches related to the objectives more frequently. Hence, to reduce the impact of semantic noise, we increase the masking probability of the patches effected severely by the semantic noise based on its statistics.
 
\subsection{Discrete Codebook for Encoded Feature Representation}
We aim to design a discrete codebook for the encoded feature space and represent encoded features by the basis vectors in the codebook. We consider important task-related features and neglect task-unrelated features with noise and imperceptible details. Specifically, we set these basis vectors as trainable parameters and train them together with the parameters of both encoder and decoder. The encoder network outputs continuous encoded features and then maps them into the discrete indices of basis vectors in the trained codebook. This design comes with the following advantages: 
\begin{itemize}
\item It is simple to train the codebook with a small variance, which makes the semantic communication system more stable.

\item The discrete feature representation can combat semantic noise.

\item The transmitter simply needs to transmit the indices of the basis vectors, which significantly reduces the transmission overhead.
\end{itemize}

\subsubsection{Codebook Design}
As in Fig. \ref{MAEStruc}(a), we denote the codebook of the encoded features as $\mathcal{E}\triangleq \big[\mathbf{e}_{1}, \mathbf{e}_{2}, \cdots, \mathbf{e}_{J} \big] \in \mathbb{R}^{J\times D}$, which consists of $J$ basis vectors $\{\mathbf{e}_{j} \in \mathbb{R}^{D}, j\in 1, 2, \cdots, J\}$ and $D$ is the dimension of each basis vector, $\mathbf{e}_{j}$. The model takes an input $\mathbf{s}$ and it passes through an encoder to produce the encoded feature vector, $\mathbf{z}_{e}(\mathbf{s})$. Then, it is mapped to a basis vector, $\mathbf{z}_{b}(\mathbf{s})$, by the nearest neighbor look-up
\begin{equation} \label{VQVAEform}
\mathbf{z}_{b}(\mathbf{s})=\textrm{arg} \min\limits_{ \mathbf{e}_{j} } \big\| \mathbf{z}_{e}(\mathbf{s}) - \mathbf{e}_{j} \big\|_{2}, \forall \mathbf{e}_{j},
\end{equation}
where the indices of features are omitted for clarity. Then, $\mathbf{z}_{b}(\mathbf{s})$ is input to the decoder. We can treat this forward computation as a layer of DNN with a particular non-linear function that maps the encoded feature vector, $\mathbf{z}_{e}(\mathbf{s})$, to a basis vector, $\mathbf{z}_{b}(\mathbf{s})$. The basis vectors, $\{\mathbf{e}_{j}, \forall j\}$, in the codebook, $\mathcal{E}$, are trained together with the parameters of encoder and decoder. 
However, operation \eqref{VQVAEform} is non-differentiable. Thus, in back propagation, we approximate the gradient by straight-through estimator \cite{VQVAE} and copy gradients from decoder input, $\mathbf{z}_{b}(\mathbf{s})$, to encoder output, $\mathbf{z}_{e}(\mathbf{s})$.  Therefore, the nearest basis vector, $\mathbf{z}_{b}(\mathbf{s})$, is passed to the decoder in forward propagation, and during the back propagation, the gradient, $\nabla_{ \mathbf{z}_{b}(\mathbf{s}) } \mathcal{L}_{c}$, is passed unaltered to the encoder. Note that the gradients contain useful information for training the encoder to minimize the loss function and can push the encoder's output, $\mathbf{z}_{e}(\mathbf{s})$, to be discretized efficiently to achieve better performance. 

\subsubsection{Differentiable Loss Function}
Our designed loss function consists of three components representing different parts of parameters: 
\begin{equation} \label{VQVAEloss}
\begin{aligned}
\mathcal{L}_{c}(\mathbf{s}, \mathbf{z}; \bm{\theta}, \mathbf{e}_{j}) &= \big\|\hat{\mathbf{s}}-\mathbf{z} \big\|_{2}^{2} + \big\| \textrm{ng}\big[ \mathbf{z}_{e}(\mathbf{s}) \big] - \mathbf{e}_{j} \big\|_{2}^{2} \\
& \quad \quad \quad \quad \quad + \beta \big\| \mathbf{z}_{e}(\mathbf{s}) - \textrm{ng}\big[ \mathbf{e}_{j} \big] \big\|_{2}^{2},
\end{aligned}
\end{equation}
where $\mathbf{s}$, $\hat{\mathbf{s}}$, and $\mathbf{z}$ denote the input, output, and true label of the network, respectively, $\bm{\theta}$ denotes the trainable parameters of the original DNN, and $\beta$ is the hyper-parameter. Symbol $\textrm{ng}\big[ \mathbf{z}_{e}(\mathbf{s}) \big]$ represents that there is no gradient passed to $\mathbf{z}_{e}(\mathbf{s})$ and its gradient is zero, which effectively constrains $\mathbf{z}_{e}(\mathbf{s})$ to be a non-updated constant. The first term is the reconstruction loss that trains the parameters of encoder and decoder. Due to the straight-through gradient estimation of mapping from $\mathbf{z}_{e}(\mathbf{s})$ to $\mathbf{z}_{b}(\mathbf{s})$, the basis vectors, $\{\mathbf{e}_{j}, \forall j\}$, receive no gradients from the reconstruction loss $\|\hat{\mathbf{s}}-\mathbf{z}\|_{2}^{2}$. Therefore, in order to train the basis vectors, we employ the $l_{2}$ error to move the basis vectors towards the encoded features, $\mathbf{z}_{e}(\mathbf{s})$, as shown in the second term of \eqref{VQVAEloss}. Since the volume of the encoded feature space is dimensionless, the codebook can grow arbitrarily and cause the training process to diverge if the basis vectors, $\{\mathbf{e}_{j}, \forall j\}$, are not trained as fast as the encoder parameters. To address this issue, we add the third term in \eqref{VQVAEloss}. In summary, the decoder is optimized by the first loss term only, the encoder is optimized by the first and the last loss terms, and the basis vectors are optimized by the middle loss term.

\subsection{Robustness of Codebook} 
\subsubsection{Semantic Similarity}
The semantic similarity of two basis vectors, $\mathbf{e}_{1}$ and $\mathbf{e}_{2}$, in the codebook can be defined as the vector multiplication, $\mathbf{e}_{1}^{T}\cdot \mathbf{e}_{2}$, cosine distance, $\dfrac{\mathbf{e}_{1}^{T}\cdot \mathbf{e}_{2}}{\|\mathbf{e}_{1}\| \cdot \|\mathbf{e}_{2}\| }$, or $l_{2}$-norm, $-\|\mathbf{e}_{1}-\mathbf{e}_{2}\|$, etc. The two basis vectors contain similar semantic information when their semantic similarity is high. We choose the cosine distance and compute all the semantic similarity between two basis vectors in the codebook. Denote the normalized codebook that consists of all the normalized basis vectors as 
\begin{equation} 
\mathbf{E}\triangleq \bigg[ \dfrac{\mathbf{e}_{1}}{ \| \mathbf{e}_{1} \| }, \dfrac{\mathbf{e}_{2}}{ \| \mathbf{e}_{2} \| }, \cdots, \dfrac{\mathbf{e}_{J}}{ \| \mathbf{e}_{J} \| } \bigg]. 
\end{equation} 
Hence, the $(i,j)$-th element of matrix $\mathbf{E}^{T}\mathbf{E}$ denotes the semantic similarity of basis vectors $\mathbf{e}_{i}$ and $\mathbf{e}_{j}$.  

\subsubsection{Improved Robust Codebook} \label{OrthogLoss}
The loss function for decreasing the semantic similarity, i.e., increasing the distance, among the basis vectors can be written as
\begin{equation} \label{Transmission}
\mathcal{L}_{s}=\|\mathbf{E}^{T}\mathbf{E}\|_{2}.
\end{equation}  
Based on semantic similarity, we add term $\|\mathbf{E}^{T}\mathbf{E}\|_{2}$ into loss function \eqref{VQVAEloss} and try to make the basis vectors in the codebook, $\mathcal{E}$, mutually orthogonal, i.e., the semantic similarity between two basis vectors is small and their distance is large.

\subsubsection{Analysis of Codebook Robustness}
The discrete representation is efficient to reduce the impact of the semantic noise. As shown in Fig. \ref{MAEStruc}(b), the semantic noise causes the extracted feature to move towards some certain directions. Only if it does not leave far away from the basis vector corresponding to the original extracted feature that is not affected by the semantic noise, the impact of the semantic noise can be eliminated by this discrete representation. Thus, increasing the distance between two basis vectors can improve the robustness of the codebook against the semantic noise. Based on the semantic similarity, we propose a novel loss function to increase the distance between two basis vectors and try to make them mutually orthogonal. The orthogonal basis vectors have two advantages: (i) the distance between two orthogonal basis vectors is large; (ii) the required number of basis vectors to represent the encoded feature space is the least when basis vectors are mutually orthogonal. 
  
\subsection{Efficient Transmission with Codebook}
\subsubsection{Codebook-Based Constellation Diagram}
Existing works in semantic communications focus on mapping the source data directly into channel symbols and assume the full-resolution constellation \cite{JSCC,DeepSC,DynamicSNR}, that is, the constellation points can appear anywhere in the constellation. However, the full-resolution constellation is extremely complex for practical systems. Thus, we need to limit the number of constellation points. To address this issue, we propose a discrete codebook-based system, which is more practical for digital communication systems, since the indices of the features can be mapped into existing finite constellation and it can better fit the digital communication scheme. In general, discretization would cause the loss of data transmission accuracy. Fortunately, VQ-VAE is proved to be an efficient vector quantization scheme that achieves satisfactory performance \cite{VQVAE}. Compared to the conventional uniform quantization, VQ-VAE achieves better quantization performance since the codebook is jointly trainied with the system.

\subsubsection{Semantic Communications with Efficient Transmission}
We assume that the transmitter and receiver share the codebook, $\mathcal{E}$, consisting of the basis vectors, $\{\mathbf{e}_{j}, \forall j\}$, which are fixed after the training stage. Thus, for each encoded feature output by the encoder, the transmitter simply needs to send the index of the corresponding basis vector, which significantly reduces the transmission overhead. As shown in Fig. \ref{MAEStruc}(a), in the transmission stage, the indices of the encoded features are firstly mapped into the binary bits. Then, these binary bits are mapped into symbols and transmitted via wireless channel, $\mathbf{H}$. The receiver maps the received symbols into the indices and finds the corresponding basis vectors in the codebook, which are the input into the decoder for further processing. The channel and noise can be treated as a layer of the autoencoder and jointly trained with the parameters of the autoencoder.    

\section{Feature Importance Module with Training Method}  \label{Efficient}
In this section, we make the semantic communication system more robust and efficient by designing the FIM. Furthermore, the SNR is incorporated into the FIM, which ensures that the proposed system can successfully operate with different SNR levels. Moreover, we propose a novel loss function and training method to train the FIM. 

\begin{figure*}[t]
\begin{centering}
\includegraphics[width=0.85\textwidth]{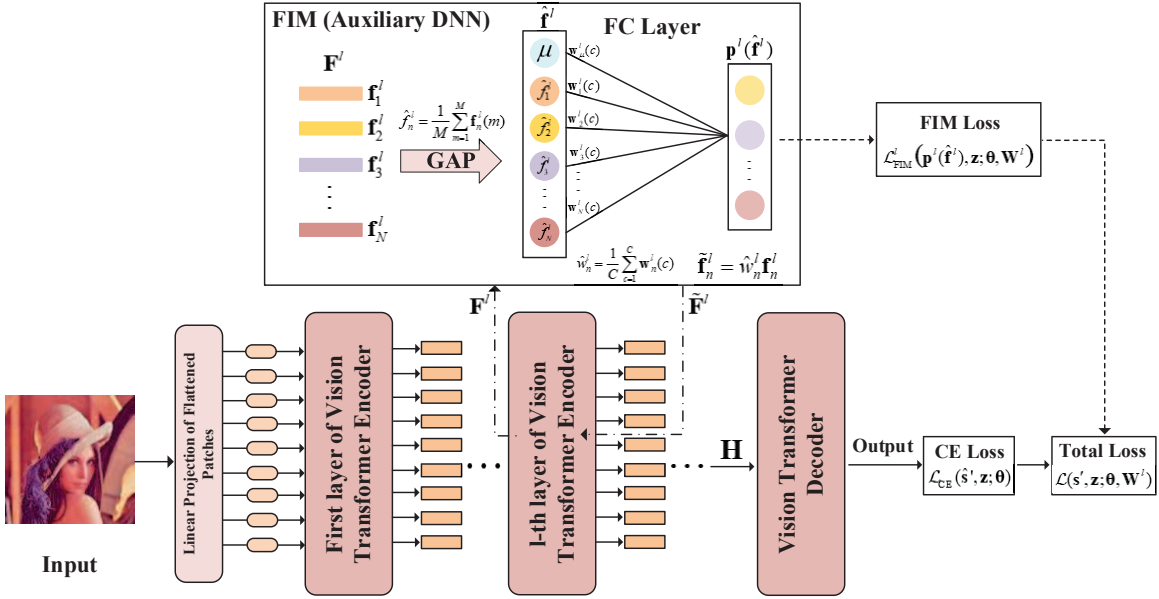}
\par\end{centering}
\caption{The architecture of the MAE with FIM.}
\label{FeatureImport}
\end{figure*}

\subsection{Noise-Related Feature Suppression}
The FIM dynamically learns and incorporates the feature importance into the training phase to train a DNN model that inherently suppresses those noise-related and task-unrelated features. 
 
\subsubsection{Noise-Related Features}
Different features depict an image from different aspects and there exist strong connections between features and robustness to semantic noise, where such robustness varies with different features. Different from the existing works assuming features are of equal importance, we focus on the relationship among features and assign them different importance. Intuitively, different features contribute differently to the results of tasks and have different levels of robustness to semantic noise. We expect that the semantic communication systems can learn the importance of different features and have better understanding of semantic information behind the input image. Then, the transmitter can send the important task-related and noise-unrelated features, which significantly improves the robustness of system and reduce the transmission overhead.

\subsubsection{Feature Activation}
We observe two characteristics of semantic noise from the feature activation perspective: (i) the magnitudes of the activated features from the samples with semantic noise are higher than that of natural samples; (ii) some noise-related features are activated more uniformly and frequently by samples with semantic noise. We find that the adversarial training has addressed the first issue of high magnitudes of activated features. In other words, some task-unrelated and low contributing features that are not activated by clean samples without semantic noise, yet are activated by samples with semantic noise. This to some extent explains why adversarial training works but its performance is unsatisfactory. It motivates us to design an FIM that trains a model to assign different importance to the features and inherently suppress these task-unrelated and noise-related high magnitude features from being activated by semantic noise. 

\subsection{FIM with Dynamic SNR}
\subsubsection{Architecture of FIM}
We denote the raw feature map, $\mathbf{F}^{l}\triangleq \big[ \mathbf{f}^{l}_{1}, \mathbf{f}^{l}_{2}, \cdots, \mathbf{f}^{l}_{N} \big]\in \mathbb{R}^{M\times N}$, as the output of the $l$-th layer, where $M$ and $N$ represent the dimension and the number of features, respectively. As shown in Fig. \ref{FeatureImport}, we first apply the global average pooling (GAP) operation on the raw feature map, $\mathbf{F}^{l}$, to obtain the feature activation map, $\hat{\mathbf{f}}^{l}\triangleq \big[\mu, \hat{f}^{l}_{1}, \hat{f}^{l}_{2}, \cdots, \hat{f}^{l}_{N} \big]\in \mathbb{R}^{N+1}$, where $\mu$ denotes the SNR. For the $n$-th feature, we have
\begin{equation} \label{GAPFeature}
\hat{f}^{l}_{n}=\frac{1}{M}\sum\limits_{m=1}^{M} \mathbf{f}^{l}_{n}(m).
\end{equation} 
Note that the GAP extracts the global feature information by averaging elements of the feature vector \cite{DynamicSNR}. Moreover, different SNR levels result in different importance of features. Thus, to ensure that the proposed semantic communication system can operate in a wide range of SNR levels, $\mu$ is designed as a part of the input of the FIM, $\hat{\mathbf{f}}^{l}$. 

The feature activation map, $\hat{\mathbf{f}}^{l}$, is then passed into an auxiliary DNN with fully-connected (FC) layers and ReLU function. The output of the semantic communication system is $\mathbf{z}\in \mathbb{R}^{C}$, e.g., class label for classification problems and reconstructed images for reconstruction problems. Then, the trainable parameters of this auxiliary DNN can be written as $\mathbf{W}^{l}\triangleq \big[\mathbf{w}^{l}_{\mu}, \mathbf{w}^{l}_{1}, \mathbf{w}^{l}_{2}, \cdots, \mathbf{w}^{l}_{N} \big]\in \mathbb{R}^{C\times (N+1)}$, which identifies the importance of each feature corresponding to output $\mathbf{z}$. Parameters $\mathbf{W}^{l}$ will be applied to reweight the raw feature map $\mathbf{F}^{l}$. We denote the weight component as $\hat{\mathbf{w}}^{l}\triangleq \big[ \hat{w}^{l}_{1}, \hat{w}^{l}_{2}, \cdots, \hat{w}^{l}_{N} \big] \in \mathbb{R}^{N}$, where we have
\begin{equation} \label{ReWeight}
\hat{w}^{l}_{n}=\frac{1}{C}\sum\limits_{c=1}^{C} \mathbf{w}^{l}_{n}(c),
\end{equation} 
for the $n$-th feature. It is employed as the weight of importance for the $n$-th feature in the $l$-th layer associated with output $\mathbf{z}$. We apply a ``softmax" layer to scale these weights into range $[0,1]$. Then, it reweights the features, $\{\mathbf{f}^{l}_{n}, \forall n\}$, in the raw feature map, $\mathbf{F}^{l}$, as $\tilde{\mathbf{f}}^{l}_{n}= \hat{w}^{l}_{n}\mathbf{f}^{l}_{n}$.
The adjusted feature map, $\tilde{\mathbf{F}}^{l}\triangleq \big[ \tilde{\mathbf{f}}^{l}_{1}, \tilde{\mathbf{f}}^{l}_{2}, \cdots, \tilde{\mathbf{f}}^{l}_{N} \big]\in \mathbb{R}^{M\times N}$, will be passed into the next layer via forward propagation. In this way, the feature relationship is captured and different weights are generated for different features to increase or suppress their connection strength to the next layer. 

\subsubsection{FIM with Label Information}
Moreover, to make full use of the label information, we slightly modify our FIM for the image classification task and incorporate the label information to improve the performance. Particularly, at the training phase, the ground-truth label is utilized as the index to determine the channel importance. While in the inference phase, we simply take the weight component that is associated to the predicted class as the feature importance since the ground-truth label is not available. Thus, the feature importance is then applied to reweight the original activation map as
$\tilde{\mathbf{f}}^{l}_{n}= \mathbf{w}^{l}_{n}(y)\mathbf{f}^{l},$ where $y$ denotes the true label at the training phase and the predicted class at the inference phase, $\tilde{\mathbf{F}}^{l}\triangleq \big[ \tilde{\mathbf{f}}^{l}_{1}, \tilde{\mathbf{f}}^{l}_{2}, \cdots, \tilde{\mathbf{f}}^{l}_{N} \big]\in \mathbb{R}^{M\times N}$. Then, the adjusted $\tilde{\mathbf{F}}^{l}$ is passed into the next layer via forward propagation. Therefore, in this case, the feature importance is learned by considering the label information. 

\subsection{Model Training}
\subsubsection{Loss Function of FIM}
We insert the proposed FIM into certain layers of the original DNN, which can be considered as an auxiliary network. It can be trained together with the parameters of the encoder and decoder by using the adversarial training method. We design the loss functions to simultaneously train the original DNN and the FIM, where by taking one inserted FIM after the $l$-th layer as an example, the loss function is designed as
\begin{equation} \label{FIMloss1}
\mathcal{L}_{ \textrm{FIM} }^{l} \big( \mathbf{p}^{l}( \hat{\mathbf{f}}^{l}), \mathbf{z}; \bm{\theta}, \mathbf{W}^{l} \big) \triangleq \mathcal{L}_{ \textrm{CE} }( \mathbf{p}^{l}(\hat{\mathbf{f}}^{l}), \mathbf{z}; \bm{\theta}, \mathbf{W}^{l} ),
\end{equation} 
where $\mathbf{p}^{l}(\hat{\mathbf{f}}^{l})\triangleq \textrm{ReLU} (\mathbf{W}^{l} \hat{\mathbf{f}}^{l})$ is the output of the FIM, $\bm{\theta}$ denotes the trainable parameter of the original DNN, $\mathbf{W}^{l}$ is the trainable parameter of the FIM, and $\mathcal{L}_{ \textrm{CE} }( \mathbf{p}^{l} (\hat{\mathbf{f}}^{l}), \mathbf{z}; \bm{\theta}, \mathbf{W}^{l} )$ is the cross entropy loss of $\mathbf{p}^{l}( \hat{\mathbf{f}}^{l})$ and label $\mathbf{z}$. 
The loss function of FIM is designed as \eqref{FIMloss1} since the FIM will be trained better, i.e., the noise-related features are suppressed and the system achieves better performance, when $\mathbf{p}^{l}(\hat{\mathbf{f}}^{l})$ approaches label $\mathbf{z}$.

It can be easily extended to multiple FIMs. The overall loss function for adversarial training with the proposed FIM can be written as
\begin{equation} \label{FIMloss2}
\!\! \mathcal{L}_{f}(\! \mathbf{s}', \mathbf{z}; \bm{\theta}, \mathbf{W}^{l} \!) \!=\! \mathcal{L}_{ \textrm{CE} }( \! \hat{\mathbf{s}}', \! \mathbf{z}; \! \bm{\theta} \!) \!+\! \frac{\gamma}{L} \! \sum\limits_{l=1}^{L} \! \mathcal{L}_{ \textrm{FIM} }^{l} \big( \! \mathbf{p}^{l}( \! \hat{\mathbf{f}}^{l} \!), \mathbf{z}; \bm{\theta}, \mathbf{W}^{l} \! \big),
\end{equation} 
where $\mathbf{s}'$ denotes the sample with semantic noise used for adversarial training, $\hat{\mathbf{s}}'$ is the output of the decoder, $\mathcal{L}_{ \textrm{CE} }( \hat{\mathbf{s}}', \mathbf{z}; \bm{\theta} )$ denotes the cross entropy loss of $\hat{\mathbf{s}}'$ and $\mathbf{z}$, $L$ denotes the number of DNN layers, and $\gamma$ is a tunable parameter for controlling the strength of FIM.  

\subsubsection{Training Method of the Whole Model}
We develop a method to jointly train the aforementioned modules in the proposed robust semantic communication system, where the detailed training procedures are summarized in Algorithm \ref{Training}. 

\begin{algorithm}[t] 
\begin{small}
\caption{Training procedures of the robust semantic communication system} 
\label{Training}
\DontPrintSemicolon
\SetKwInOut{Input}{Input}
\SetKwInOut{Output}{Output}
\SetKwInOut{Initialize}{Initialize}
\Input{The training dataset that consists of the input images, $\mathbf{s}_{i}$, with their true labels, $\mathbf{z}_{i}$, the generated channel samples, $\mathbf{H}_{i}$, and the number of training epoch, $N$. }
\Output{The discrete codebook $\mathcal{E}$ and the trained robust model with FIM, encoder, and decoder.}
\For{$n\leftarrow 1$ \KwTo $N$}{
Mask a portion of the input images based on the proposed masking strategy.\\
Calculate $\mathcal{L}_{c}$ based on the loss function \eqref{VQVAEloss} for training the discrete codebook $\mathcal{E}$, together with the encoder and decoder.\\
Calculate $\mathcal{L}_f$ based on the loss function \eqref{FIMloss2} for training the FIM.\\
Calculate $\mathcal{L}_{s}$ based on the loss function \eqref{Transmission} for fine-tune the encoder and decoder.\\
Execute Algorithm \ref{minmax} to do adversarial training via employing the sum of $\mathcal{L}_{c}$, $\mathcal{L}_{f}$, and $\mathcal{L}_{s}$. \\
}
\end{small}
\end{algorithm}

\section{Simulation Results} \label{Simulation}
In this section, we verify the effectiveness of the proposed semantic communication system by numerical results.

\begin{table} 
\centering
\caption{The architecture of the proposed model.}  
\label{NNA}
\begin{tabular}{|c|c|c|c|}
\hline
~& Layer Name & Dimension & Activation \\
\hline
\multirow{4}*{Transmitter} & 8$\times$Transformer Encoder & 768 (12 heads) & Linear \\
\cline{2-4}
& Dense & 256 & Sigmoid \\
\cline{2-4}
& Codebook & 128 & None \\
\cline{2-4}
& FIM & 196 &  ReLU \\
\hline
Channel& Channel & $N_{r}$ & None \\
\hline
\multirow{5}*{Receiver} & FIM & 196 &  ReLU \\
\cline{2-4}
& Codebook & 128 & None \\
\cline{2-4}
& Dense & 768 & Sigmoid \\
\cline{2-4}
&  4$\times$Transformer Encoder & 768 (12 heads) & Linear\\
\cline{2-4}
& Dense & 10 & ReLU \\	 
\hline  
\end{tabular} 
\end{table}

\subsection{Simulation Setup}
We consider the scenario where the transmitter and the receiver are equipped with $N_{t}=4$ transmit and $N_{r}=2$ receive antennas, respectively. We compare the proposed masked VQ-VAE with the conventional source coding and channel coding approaches under MIMO channels \cite{ConvenCom}. We adopt CIFAR-10, consisting of $60,000$ images of $10$ classes as the dataset for image classification, Cars196, with $16,185$ images of $196$ classes as the dataset for image retrieval, and ImageNet, consisting of $14,197,122$ images as the dataset for image reconstruction. The average size of joint photographic experts group (JPEG) images in the dataset is $5,108$ bytes and the number of patches of each image is $14\times 14$. The masking ratio of masked VQ-VAE is $0.5$ and the codebook size is $256$, which requires $8$ bits for transmitting each index. We adopt the 16-QAM for modulation with low-density parity-check code (LDPC) of rate $1/2$.
The number of iterations to generate semantic noise is set as $K=5$ and its power is $\epsilon=0.016$. 
The architecture of the proposed model is presented in Table \ref{NNA}. In particular, the ``Codebook" layer represents that the basis vectors in the codebook are set as trainable parameters and the ``Dimension" of ``Codebook" layer denotes the number of basis vectors. Moreover, the ``Dimension" of other layers represent the output dimension of this layer. 
For different downstream tasks, e.g., classification, we simply need to change the dimension of the last layer and fine-tune the pre-trained masked VQ-VAE. We compare the performance of the following methods:
\begin{itemize}
\item Masked VQ-VAE+FIM+AT: The proposed masked VQ-VAE with FIM and adversarial training.

\item Masked VQ-VAE+AT: The proposed masked VQ-VAE with adversarial training. 

\item Masked VQ-VAE: The proposed masked VQ-VAE.

\item JSCC+AT: The modified JSCC scheme in \cite{JSCC} for different tasks with the ViT architecture and adversarial training.

\item JSCC: The modified JSCC scheme in \cite{JSCC} for different tasks with the ViT architecture.

\item JPEG+LDPC+AT: The conventional scheme that adopts JPEG for the image source coding, LDPC for the channel coding, and the ViT as classifier with the adversarial training.

\item JPEG+LDPC: The conventional scheme with JPEG and LDPC. 
\end{itemize}

Note that we have proposed two kinds of semantic noise models: (i) sample-dependent semantic noise added at the transmitter; (ii) sample-independent semantic noise added at the receiver, where (i) has a more serious impact on semantic communication systems. Hence, unless otherwise stated, we adopt semantic noise model (i). We consider the following tasks and generate the corresponding semantic noise.
\begin{itemize}
\item For image classification, the semantic noise is generated to misclassify the minimally perturbed data that looks visually similar to clean samples.

\item As for image retrieval, the subtle semantic noise leads to incorrect retrieved results.

\item For image reconstruction, the semantic noise results in the failure of reconstruction, e.g., some key objects and information in the reconstructed images are missed or blurred. 
\end{itemize}

\subsection{Analysis of Transmission Overhead}

\begin{table*}
\centering  
\caption{The number of transmitted symbols for one image in different tasks.}    
\label{Overhead} 
\begin{tabular}{|c|c|c|c|c|}
\hline
Schemes & JPEG+LDPC & Masked VQ-VAE (Patch $=16$)   & Masked VQ-VAE (Patch $=8$)   \\ \hline
Image classification  & $53,760$ & $196$ & $784$   \\ \hline
Image retrieval       & $249,900$ & $490$ & $1,960$     \\ \hline
Image reconstruction  & $143,350$ & $490$ & $1,960$     \\ \hline
\end{tabular}
\end{table*}

Table \ref{Overhead} presents the transmission overhead of the conventional JPEG+LDPC and our proposed masked VQ-VAE. Note that ``Patch $16$" denotes the proposed masked VQ-VAE scheme with image compression ratio $16$ in the image pre-processing stage, i.e., a $16\times 16\times 3$ patch is compressed into a scalar. Hence, a larger value represents a higher compression ratio and lower transmission overhead. 
The number of transmitted symbols of the JPEG+LDPC can be computed as: $\dfrac{LHC_{n}P_{b}R_{c}}{C_{r}B_{s}}$, where $L$ and $H$ denote the length and the width of the image, respectively, $C_{n}$ is the number of channels, $P_{b}$ denotes the required number of bits for each pixel, $R_{c}$ denotes the code rate, $C_{r}$ is the compression ratio, and $B_{s}$ denotes the required bits for a symbol that depends on the modulation mode. We take the image classification as an example: $\dfrac{224\times 224\times 3\times 8\times 2}{11.2\times 4}=53,760$ symbols/image. 
Moreover, the number of transmitted symbols for the proposed masked VQ-VAE can be computed as: $\dfrac{LH J_{c}M_{r} }{P_{a}^{2} B_{s}}$, where $J_{c}$ denotes the required number of bits for transmitting an index of basis vector in the codebook, $M_{r}$ is the masking ratio, $P_{a}$ is the patch size. We take the masked VQ-VAE (Patch $=16$) for image classification as an example: $\dfrac{224\times 224\times 8\times 0.5}{16^{2}\times 4}=196$ symbols/image. Therefore, the proposed masked VQ-VAE simply requires $196/ 53760=0.36\%$ transmitted symbols of the conventional JPEG+LDPC.

\subsection{Accuracy of Image Classification}
\begin{figure*}[!t]
\centering
\subfloat[The ablation experiments.]{\centering \scalebox{0.55}{\includegraphics{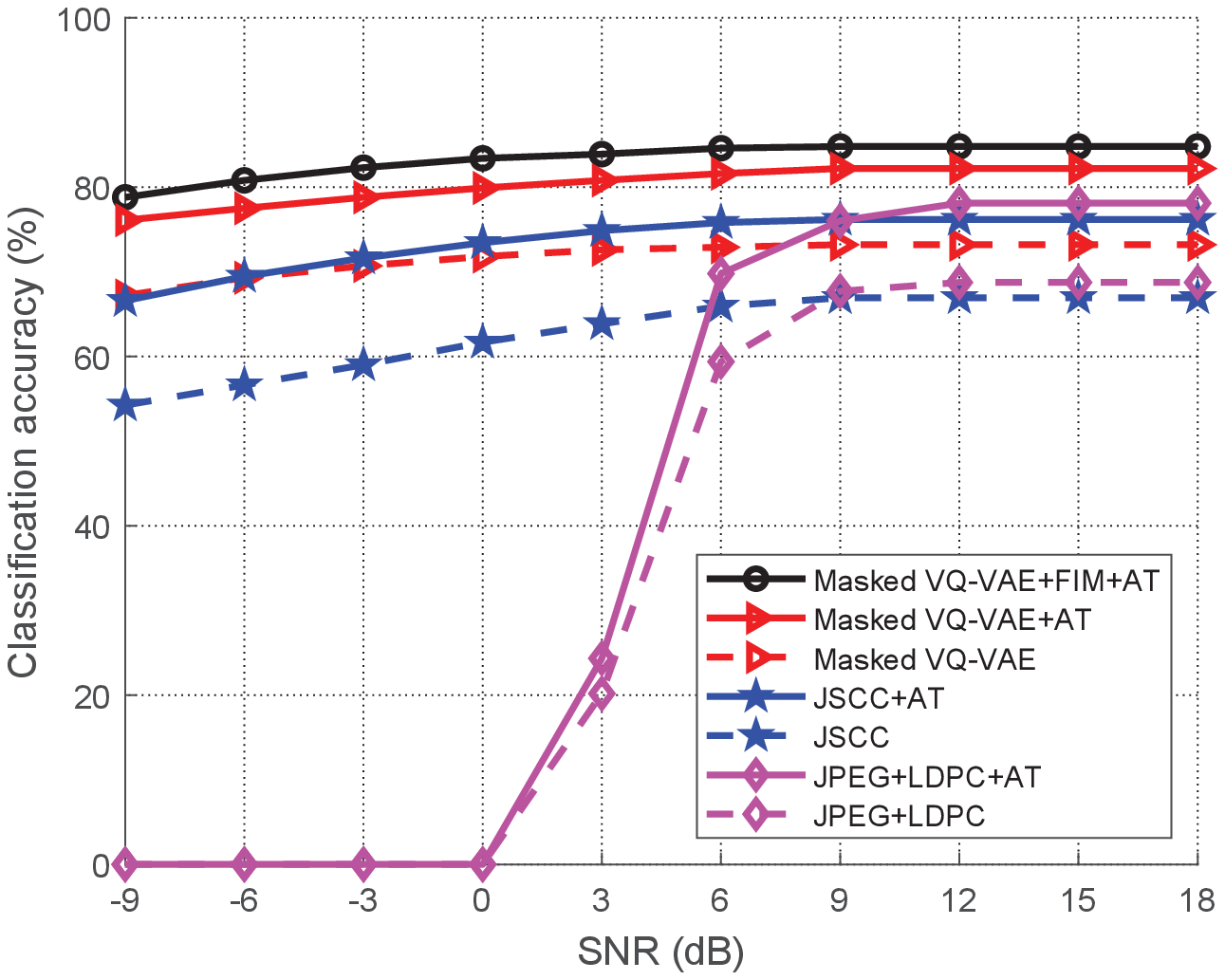}} }
\subfloat[The classification accuracy without semantic noise.]{\centering \scalebox{0.55}{\includegraphics{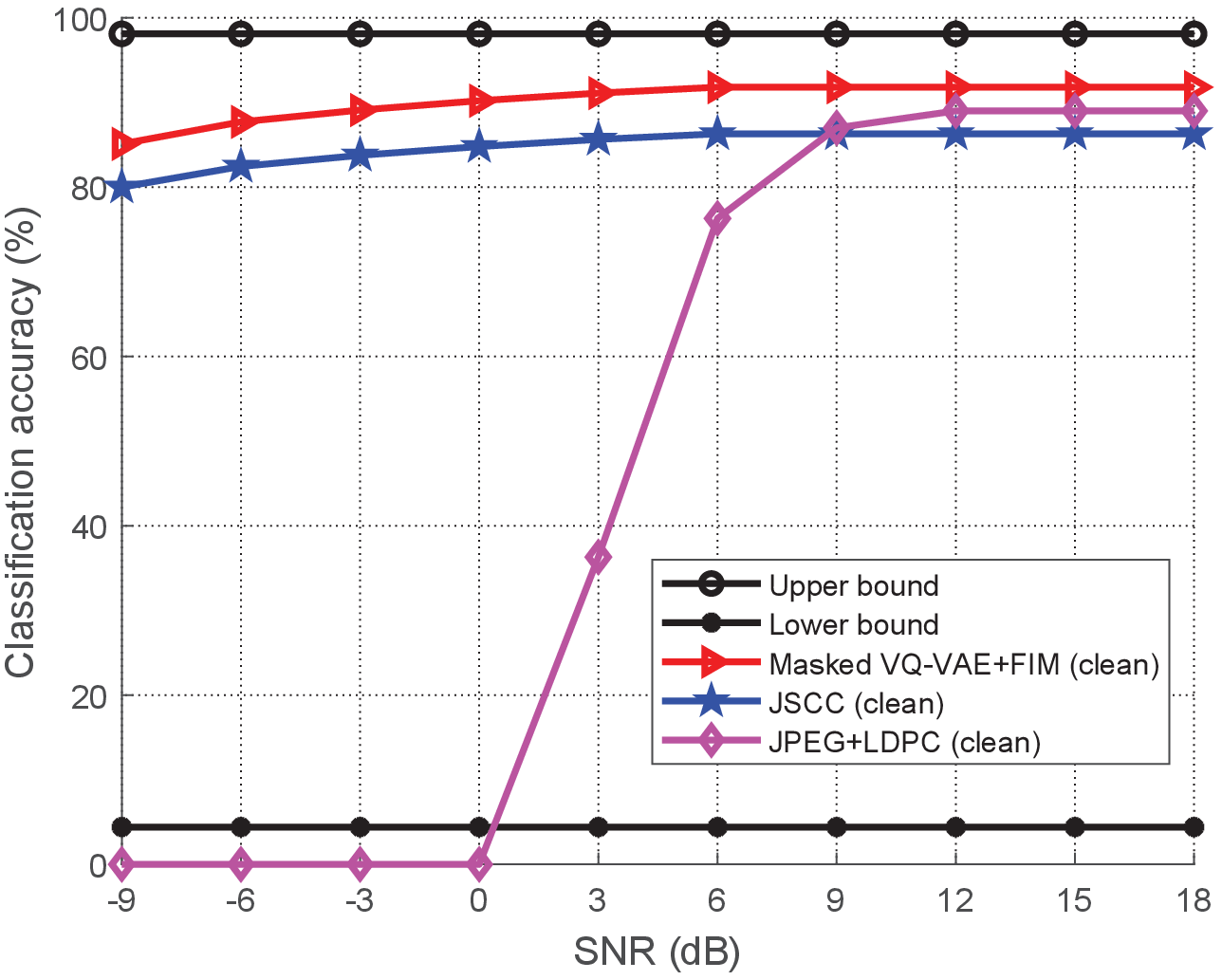}}}
\caption{The classification accuracy of the proposed schemes versus SNR. }
\label{Classification}
\end{figure*}

Fig. \ref{Classification} shows the classification accuracy versus SNR. We train the model with the SNR range from $-3$ dB to $12$ dB and test it in the SNR range from $-9$ dB to $18$ dB. From Fig. \ref{Classification}(a), the classification accuracy increases with SNR for all schemes. The proposed Masked VQ-VAE+FIM+AT significantly outperforms the JSCC+AT, JSCC, JPEG+LDPC+AT, and JPEG+LDPC, and achieves the best performance. Moreover, the proposed Masked VQ-VAE+FIM+AT outperforms the Masked VQ-VAE+AT and Masked VQ-VAE. It demonstrates the efficiency of each module in our design, including the masked VQ-VAE, FIM, and adversarial training. In addition, the proposed scheme and JSCC significantly outperform the conventional JPEG+LDPC in low SNR scenario because the BER is high in low SNR scenario while the proposed scheme is robust by transmitting the indices of extracted task-related features in the trained codebook. 

Fig. \ref{Classification}(b) shows the classification accuracy of the schemes without semantic noise. Note that ``clean" denotes the scheme without semantic noise. The upper bound is achieved by the JSCC without the effects of channel noise and semantic noise while the lower bound is achieved by the JSCC with semantic noise. From the figure, the proposed Masked VQ-VAE+FIM (clean) approaches the upper bound and significantly outperforms JSCC (clean) and JPEG+LDPC (clean). Moreover, the proposed Masked VQ-VAE+FIM+AT approaches the performance of the model without semantic noise, which demonstrates that the proposed model can effectively improve the system robustness by reducing the impacts of semantic noise. 

\begin{figure*}[!t]
\centering
\subfloat[Without semantic noise.]{\centering \scalebox{0.55}{\includegraphics{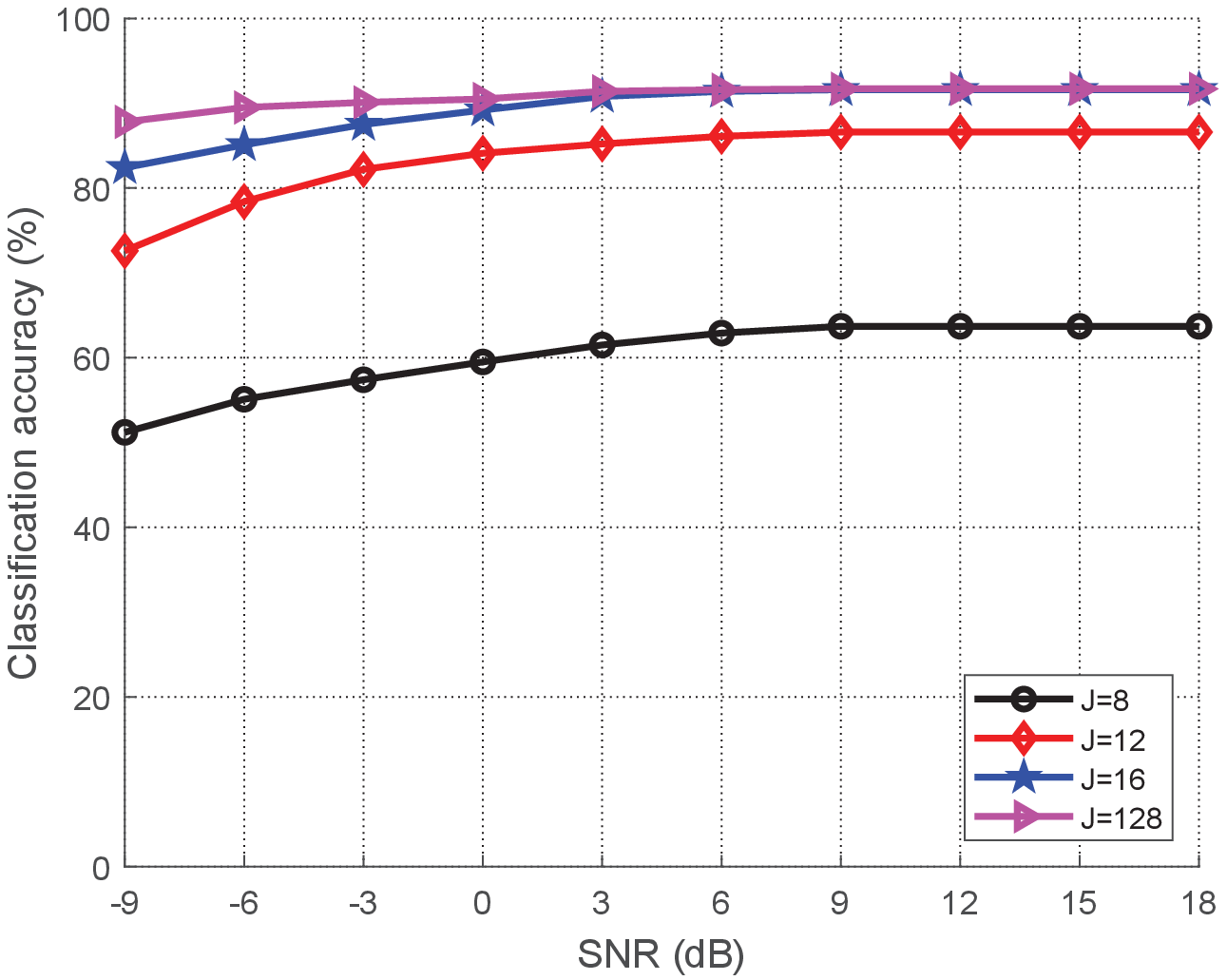}} }
\subfloat[With semantic noise.]{\centering \scalebox{0.55}{\includegraphics{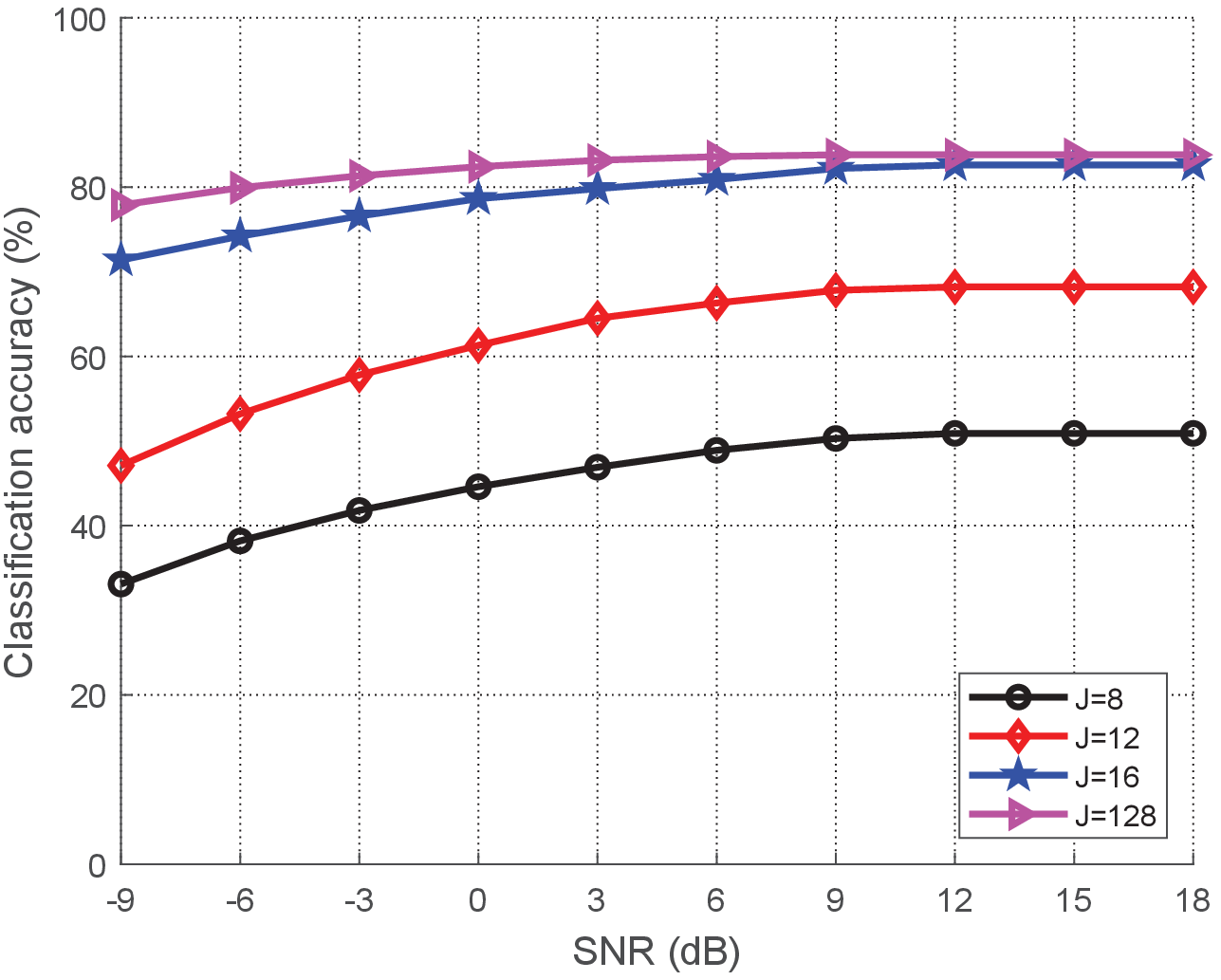}}}
\caption{The classification accuracy of different codebook size $J$ versus SNR. }
\label{Codebook}
\end{figure*}

Fig. \ref{Codebook}(a) and Fig. \ref{Codebook}(b) present the classification accuracy versus SNR for different codebook sizes without semantic noise and with semantic noise, respectively. From the figure, the classification accuracy increases with SNR. A larger codebook size has stronger robustness against semantic noise and stronger representational capability for the whole dataset, which results in a higher classification accuracy. Moreover, the codebook size $J\geq 16$ is enough for accurate classification. 

\begin{figure}[t]
\begin{centering}
\includegraphics[width=0.45\textwidth]{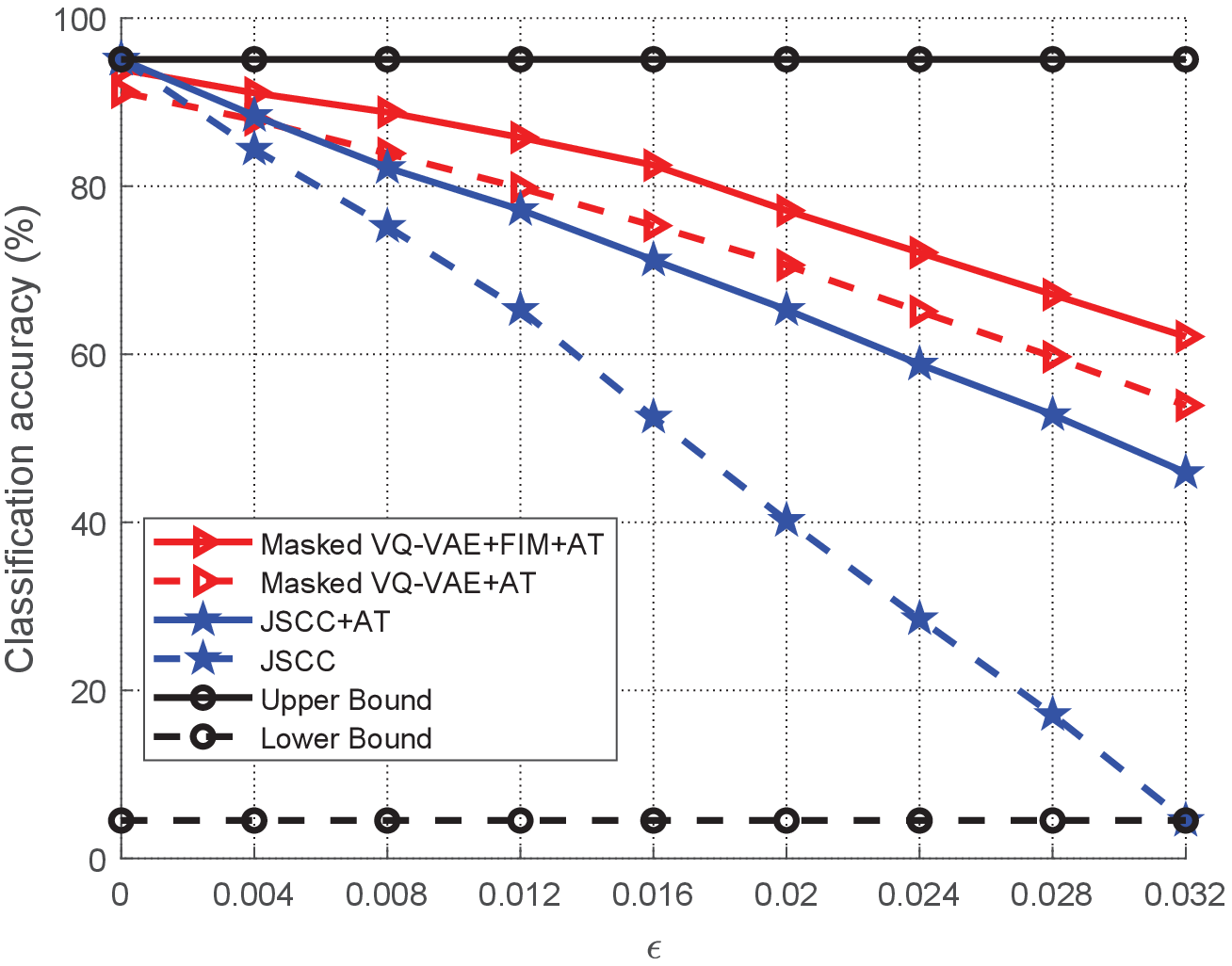}
\par\end{centering}
\caption{The classification accuracy of different schemes versus the power of semantic noise $\epsilon$.}
\label{SemPower}
\end{figure}

Fig. \ref{SemPower} illustrates the classification accuracy versus the power of semantic noise $\epsilon$. The upper bound and lower bound are achieved by JSCC without semantic noise and with maximum power of semantic noise, respectively. From the figure, the classification accuracy achieved by all schemes decreases with $\epsilon$. The proposed Masked VQ-VAE+FIM+AT significantly outperforms the benchmarks and achieves the best performance, especially when $\epsilon$ is large, which shows the superiority of the proposed model against semantic noise.

\begin{figure}[t]
\begin{centering}
\includegraphics[width=0.45\textwidth]{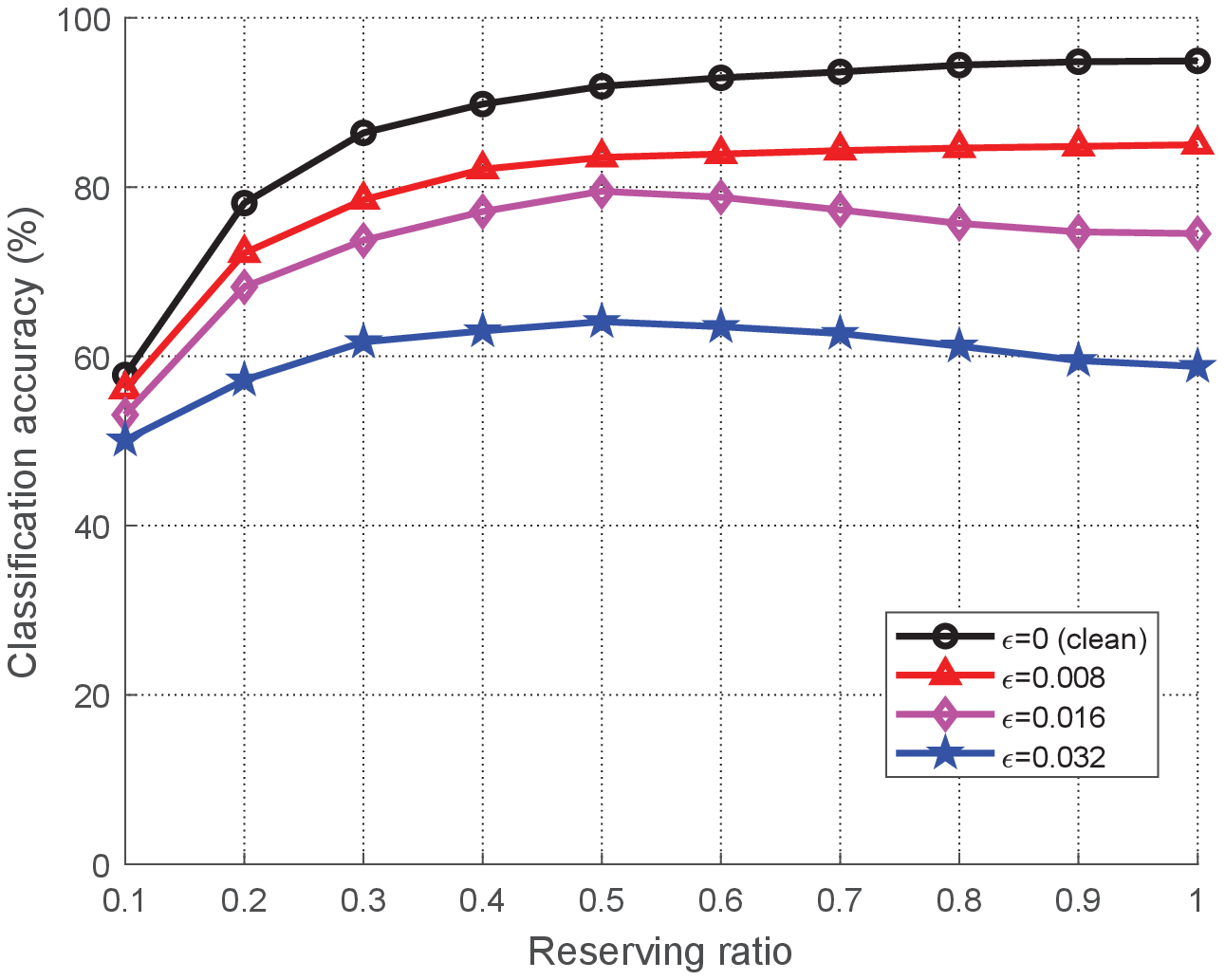}
\par\end{centering}
\caption{The classification accuracy of the proposed scheme versus reserving ratio for different $\epsilon$.}
\label{MaskRatio}
\end{figure}

Fig. \ref{MaskRatio} presents the classification accuracy of the proposed Masked VQ-VAE+FIM+AT versus reserving ratio for different $\epsilon$. Note that reserving ratio $=1-$ masking ratio, which denotes the ratio of the reserved parts of an image to that of the whole image. From the figure, the classification accuracy achieved by the proposed Masked VQ-VAE+FIM+AT decreases with $\epsilon$. In addition, when $\epsilon$ is small, e.g., $\epsilon=0.008$, the classification accuracy of the proposed scheme increases with the reserving ratio since a higher reserving ratio keeps more semantic information of the image. When $\epsilon$ is large, e.g., $\epsilon=0.016$, the classification accuracy of the proposed scheme firstly increases and then decreases with the increase of reserving ratio. It is because that a larger reserving ratio retains more semantic noise although it keeps more semantic information. Thus, reserving ratio $0.5$ achieves a good trade-off between the semantic information and semantic noise, which has the highest classification accuracy.

\begin{figure}[t]
\begin{centering}
\includegraphics[width=0.45\textwidth]{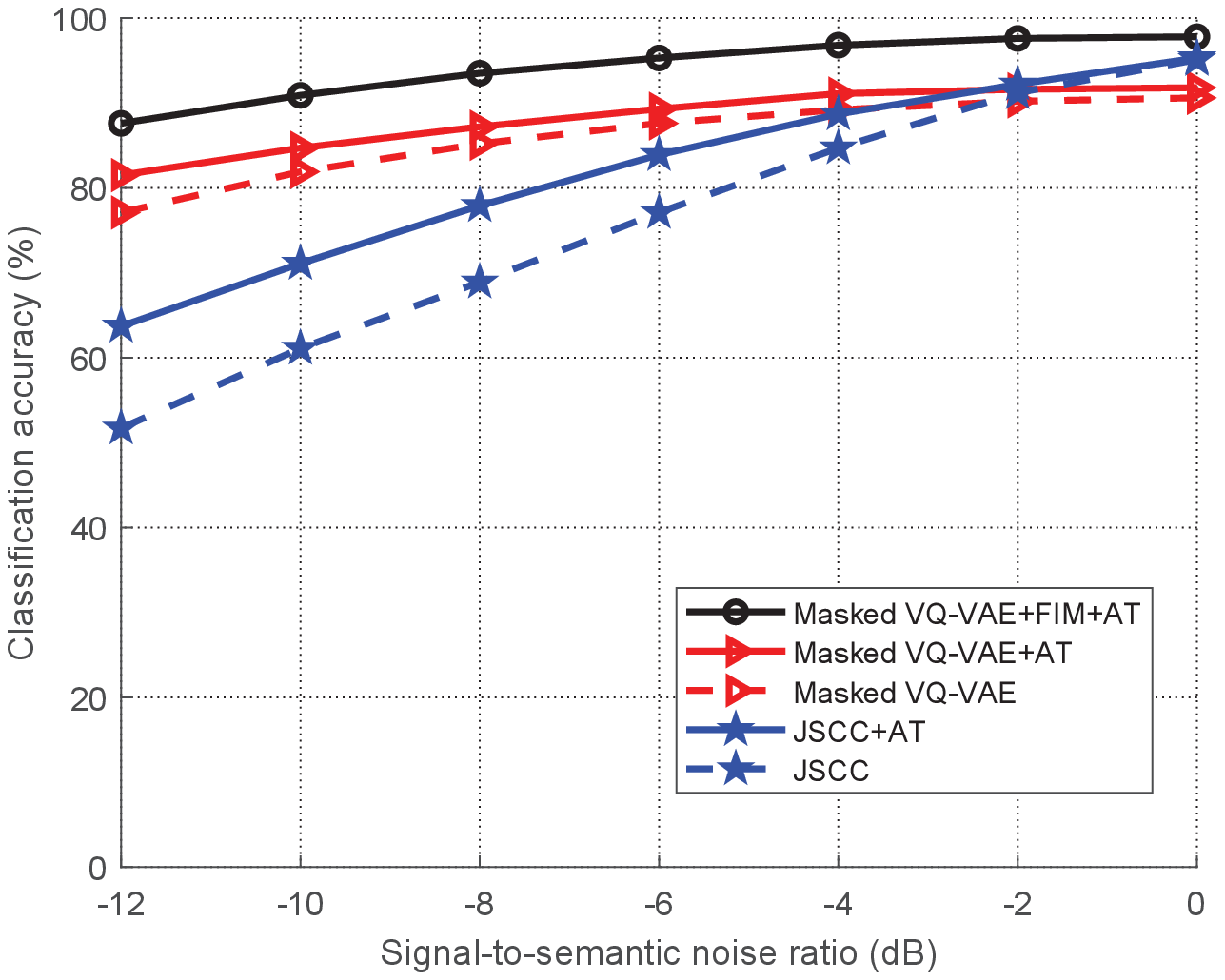}
\par\end{centering}
\caption{The classification accuracy versus signal-to-semantic noise ratio at the receiver.}
\label{SampleIndep}
\end{figure}

Fig. \ref{SampleIndep} shows the classification accuracy of different schemes versus signal-to-semantic noise ratio, where the noise denotes the sample-independent semantic noise added at the receiver. From the figure, the classification accuracy achieved by all schemes increases with signal-to-semantic noise ratio. The proposed Masked VQ-VAE+FIM+AT significantly outperforms the benchmarks and achieves the best performance. It shows that the proposed model is more robust to the sample-independent semantic noise than the JSCC. Compared with the sample-dependent semantic noise, the sample-independent semantic noise has less impacts on classification accuracy and the effects of adversarial training to combat it is not obvious.

\subsection{White-Box Semantic Noise and Black-Box Semantic Noise} 

We have claimed that the attacker needs to know all the model information to generate the semantic noise. In fact, it is almost the worst case for the proposed semantic communication system, in which the generated semantic noise can significantly influence the system performance. We call it the white-box semantic noise. However, semantic noise could also be generated when the attackers only know part of the model information, e.g., model parameters and architectures. They can also attack the system by adopting the semantic noise generated from other similar models, which is named as the black-box semantic noise. 

\begin{table*}
\centering
\caption{The classification accuracy of the white-box and black-box semantic noise.}  
\label{black-box}
\begin{tabular}{|c|c|c|c|c|c|c|c|} 
\hline
Attacked model & \multicolumn{3}{c|}{Proposed} & \multicolumn{2}{c|}{Transformer} & \multicolumn{2}{c|}{ResNet}  \\ 
\hline
Type of semantic noise   & P-P  & T-P  & R-P & T-T & R-T & R-R &T-R     \\ 
\hline
Classification accuracy & $68.7\%$ & $82.3\%$ & $85.3\%$  & $9.5\%$ & $ 75.3\%$ & $8.2\%$ & $ 67.2\%$ \\
\hline
\end{tabular}  
\end{table*}

In Table \ref{black-box}, we first generate the semantic noise for the proposed model, the Transformer-based model, and the ResNet-based model. Then, we employ these different kinds of semantic noise to evaluate the robustness of these models. Particularly, the ``T-P" in the line of ``type of semantic noise" refers to employing the semantic noise generated from the Transformer-based model to attack the proposed model, which is the black-box semantic noise. In comparison, the ``P-P" means that the proposed model is attacked with semantic noise generated from the proposed model, which is the white-box semantic noise. We can see that the white-box semantic noise has more severe effects on the performance of the proposed model than the black-box noise. Moreover, the black-box semantic noise generated from the Transformer-based model is more effective than that of the ResNet-based model since the proposed model is designed based on Transformer. It demonstrates that the black-box semantic noise generated from a more similar model tends to be more effective. It is mainly because that the models with similar architecture generally have similar parameters and layer structures. Furthermore, the proposed model is more robust than the Transformer-based and ResNet-based models against both the white-box and the black-box semantic noise. In addition, the target of this paper is to design a robust semantic communication system against semantic noise. From Table \ref{black-box}, the proposed model will be able to defense the black-box semantic noise. 

\subsection{Feature Activation}

\begin{figure*}[!t]
\centering
\subfloat[Without FIM.]{\centering \scalebox{0.31}{\includegraphics{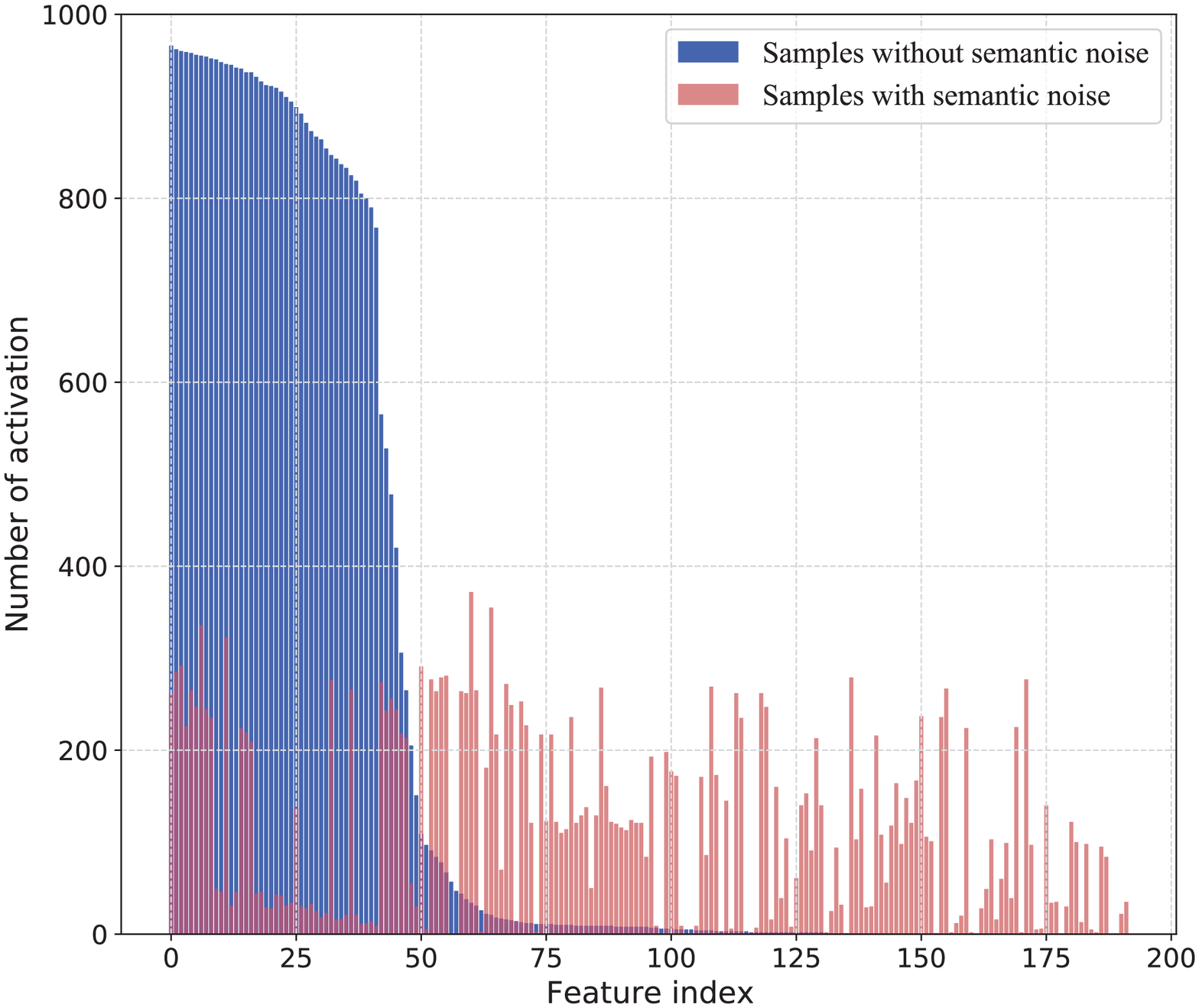}} }
\subfloat[With FIM.]{\centering \scalebox{0.31}{\includegraphics{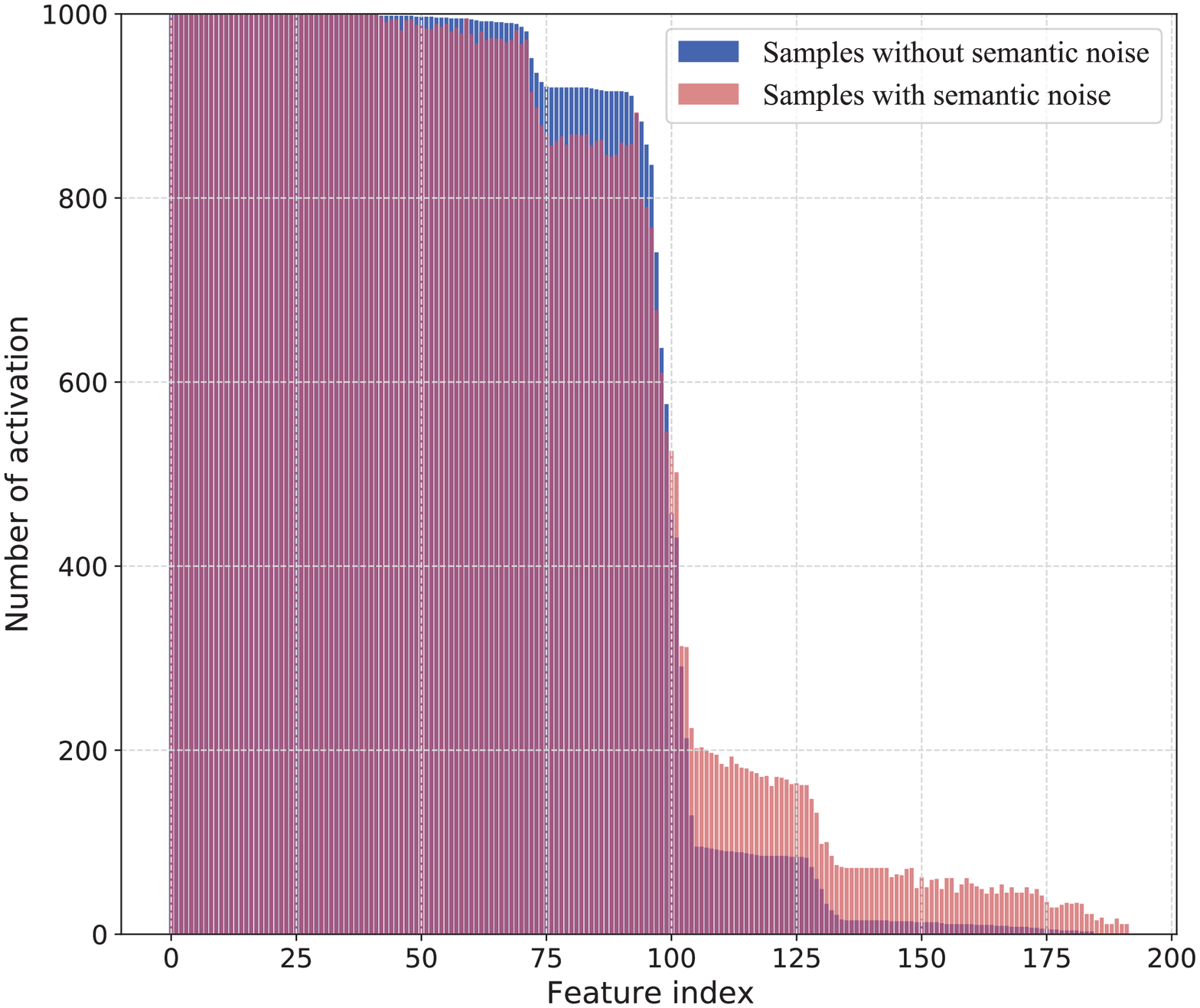}}}
\caption{The activation frequency of features. }
\label{Activation}
\end{figure*}

Fig. \ref{Activation}(a) and Fig. \ref{Activation}(b) present the activation frequency of features without FIM and with FIM, respectively. A feature is determined as activated if its activation value is larger than a threshold. 
From Fig. \ref{Activation}(a), the samples with semantic noise activate the features more uniformly and they tend to frequently activate the features that are rarely activated by clean samples, i.e., the features $125$-$200$. These low frequency features are noise-related and correspond to the redundant activation that is less important for the task. Furthermore, the semantic noise inhibits the activation of the features that are frequently activated by clean samples, i.e., the features $1$-$75$. 
Fig. \ref{Activation}(b) illustrates that the FIM increases the activation frequency of some important features, i.e., the features $75$-$100$, and avoids those redundant and less important features from being activated by samples with semantic noise, i.e., the features $125$-$200$. This result demonstrates the effectiveness of our proposed FIM, which can suppress the activation of less important and noise-related features.

\subsection{Semantic Similarity of Basis Vectors}

\begin{figure*}[!t]
\centering
\subfloat[Without semantic similarity term in loss function.]{\centering \scalebox{0.55}{\includegraphics{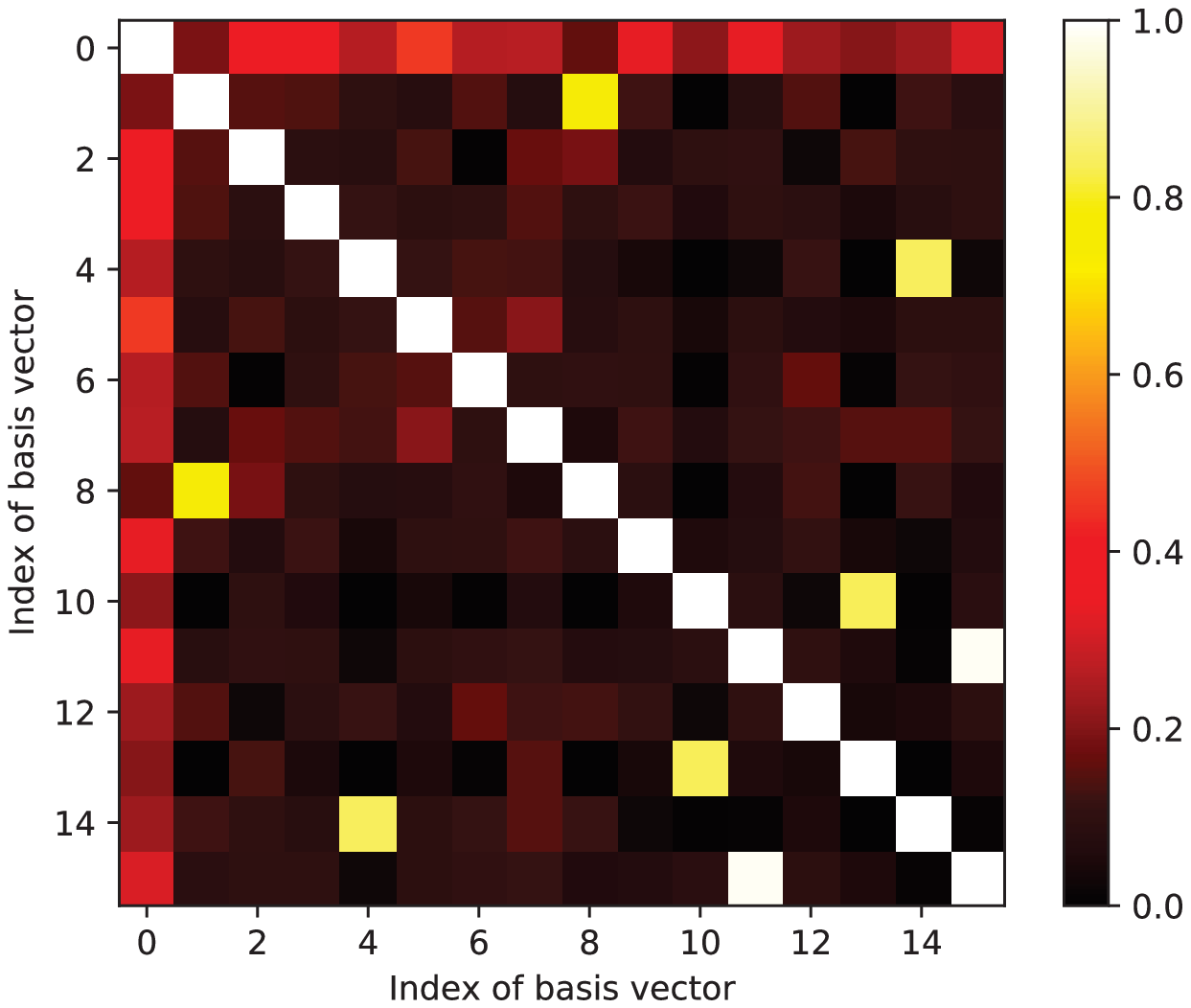}} }
\subfloat[With semantic similarity term in loss function.]{\centering \scalebox{0.55}{\includegraphics{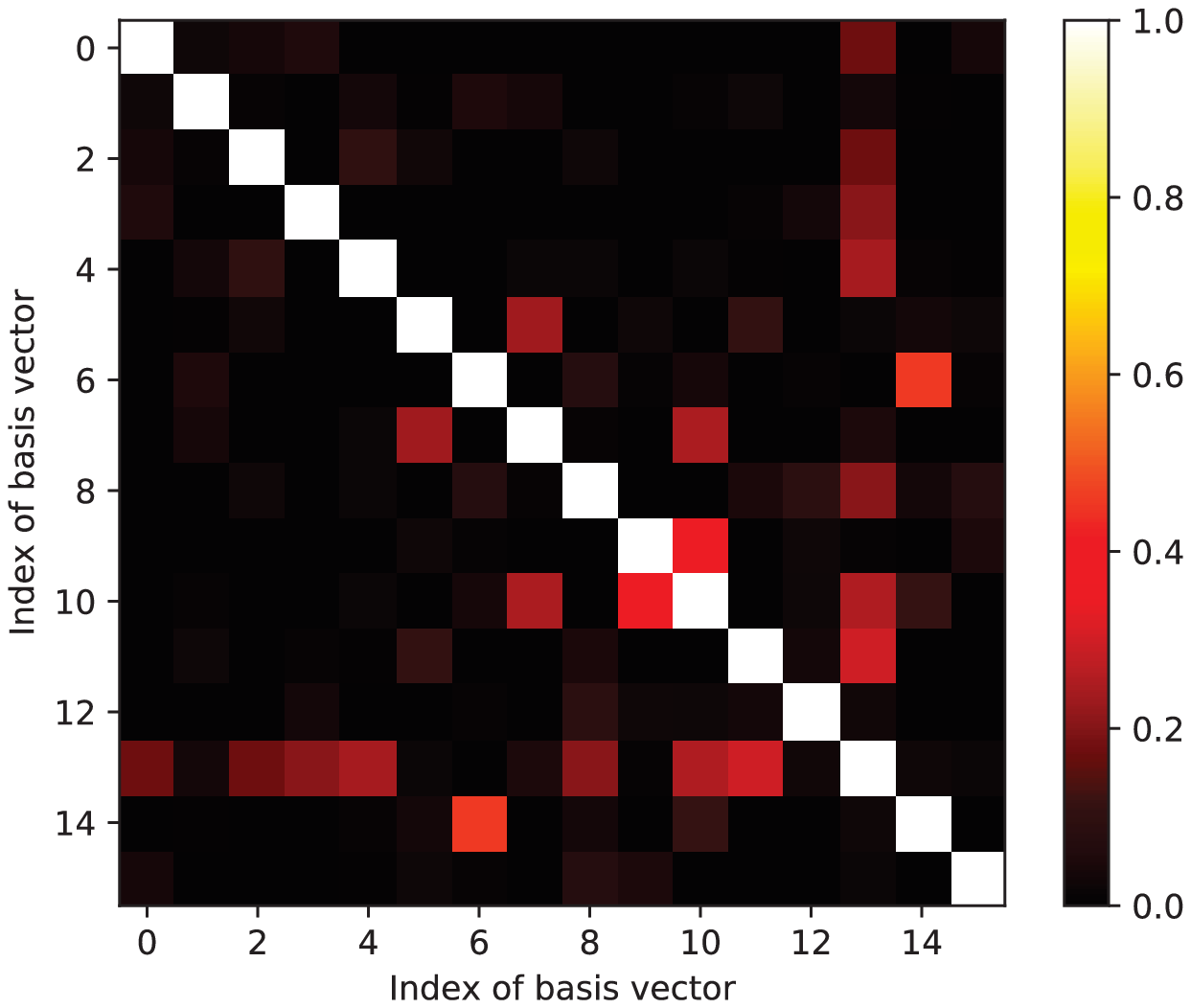}}}
\caption{Semantic similarity of basis vectors in the codebook.}
\label{Similarity}
\end{figure*}

Fig. \ref{Similarity}(a) and Fig. \ref{Similarity}(b) show the semantic similarity of basis vectors in the codebook without and with the semantic similarity term in the loss function, respectively. A larger value with brighter color represents the higher semantic similarity between the two vectors. It can be observed that semantic similarity between two basis vectors in Fig. \ref{Similarity}(a) is larger than that of Fig. \ref{Similarity}(b). It demonstrates that the semantic similarity term of loss function proposed in Section \ref{OrthogLoss}, $\|\mathbf{E}^{T}\mathbf{E}\|_{2}$, can make the basis vectors in the codebook nearly mutually orthogonal, which enhances the robustness and representational capability of the codebook. 

\subsection{Performance of Image Retrieval and Reconstruction}

\begin{figure*}[!t]
\centering
\subfloat[The recall@1 performance of image retrieval.]{\centering \scalebox{0.55}{\includegraphics{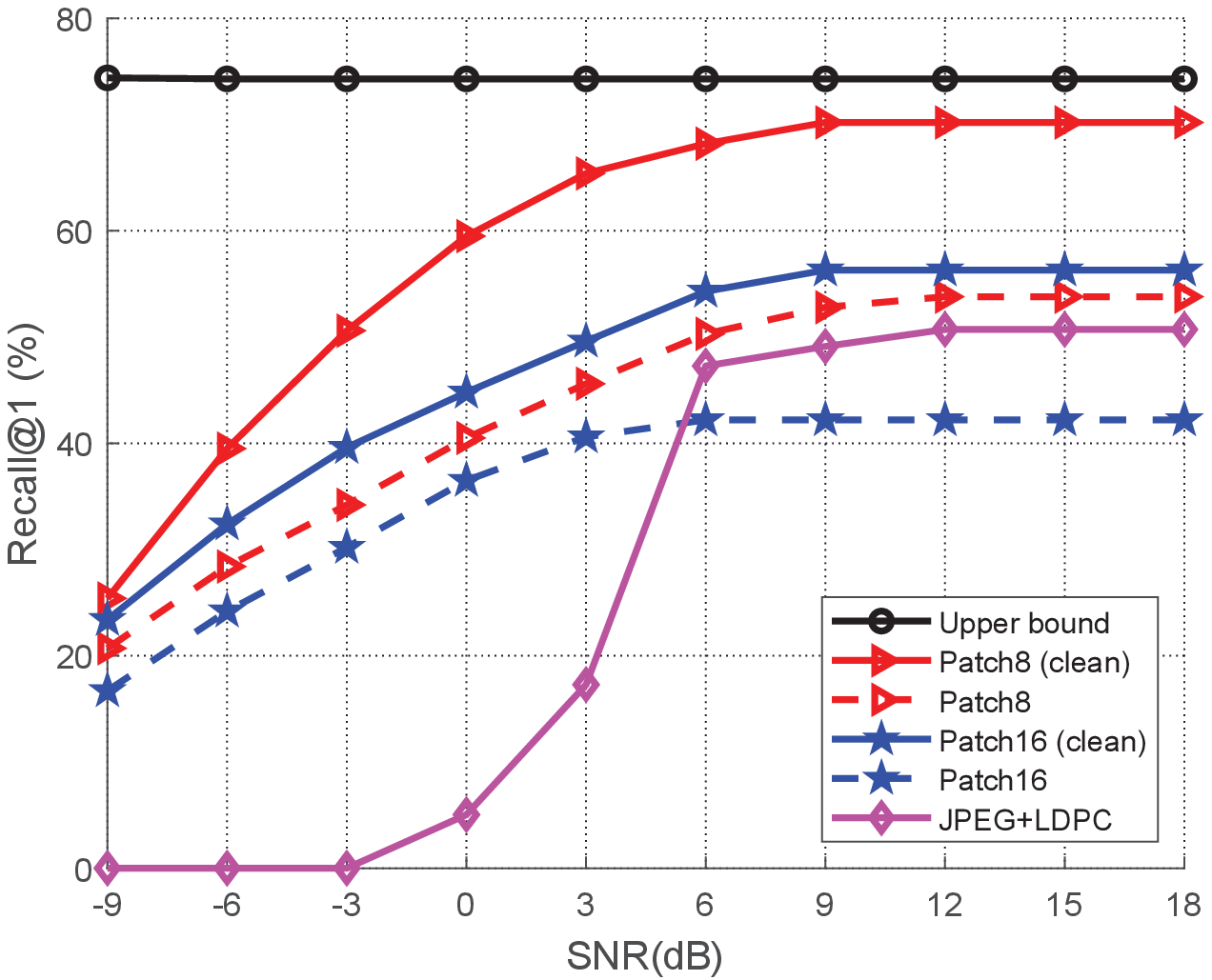}} }
\subfloat[The quality of image reconstruction.]{\centering \scalebox{0.18}{\includegraphics{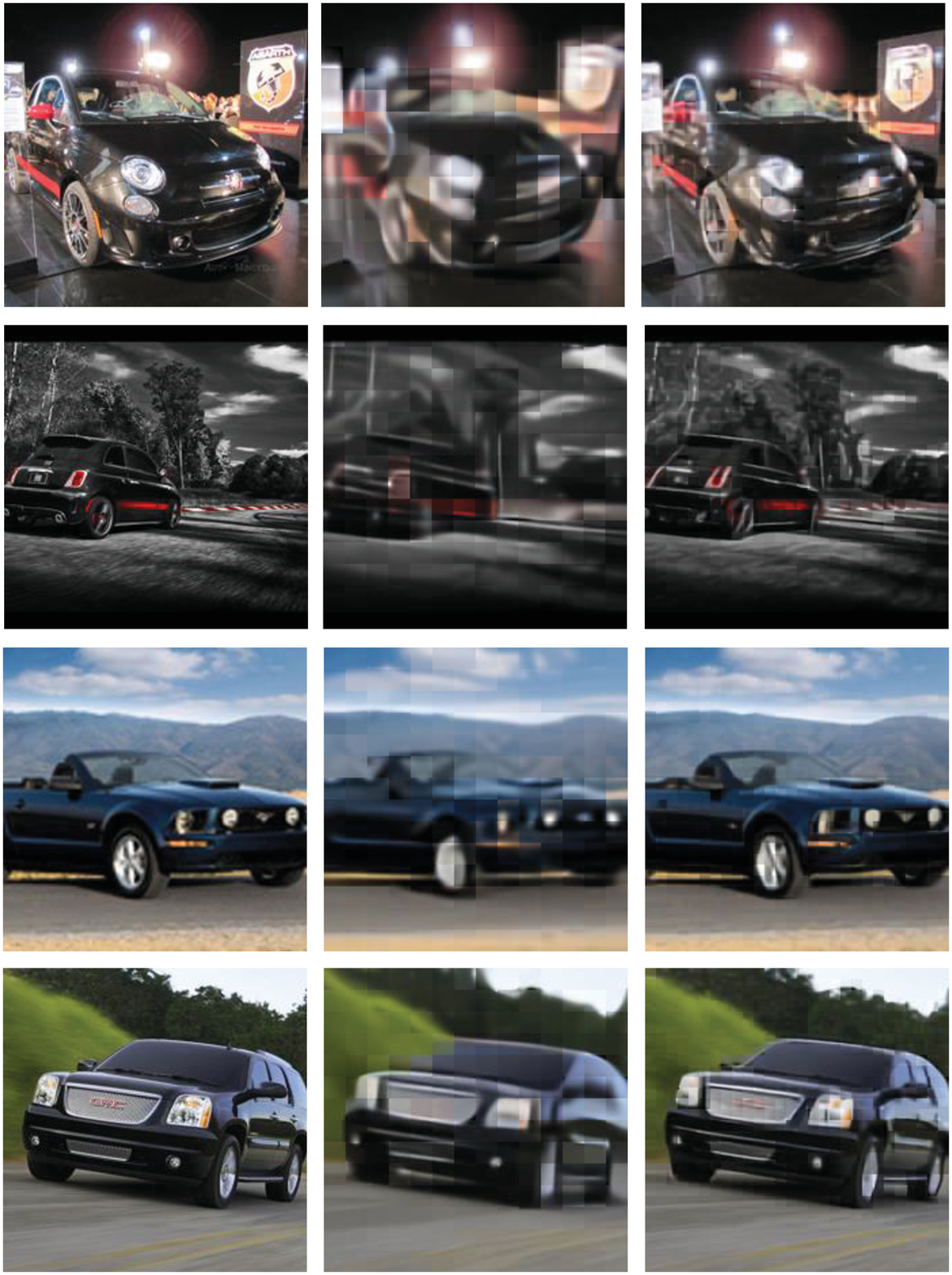}}}
\caption{The performance of image retrieval and reconstruction. }
\label{Retrieval}
\end{figure*}

Fig. \ref{Retrieval}(a) shows the recall@1 performance of image retrieval task versus SNR, where recall@1 represents the ratio of successful image retrieval at the first query. The upper bound is achieved by JPEG+LDPC without semantic noise in high SNR scenario, nearly lossless transmission. From the figure, Patch $8$ outperforms Patch $16$ that has much higher compression ratio and requires lower transmission overhead. Moreover, Patch 8 significantly outperforms the conventional JPEG+LDPC, especially in low SNR scenario. It is because that the image cannot be correctly decoded by employing JPEG+LDPC in low SNR scenario and the proposed scheme is more robust against semantic noise.  

Fig. \ref{Retrieval}(b) shows the quality of image reconstruction under the power of semantic noise $\epsilon=0.012$. The left, middle, and right column represent the original image, the image reconstructed with patch size $16$, and the image reconstructed with patch size $8$, respectively. From the figure, a lower compression ratio leads to a better construction quality and patch size $8$ achieves a satisfactory construction quality against semantic noise. 

\section{Conclusion} \label{Conclusion}
In this paper, we have analyzed semantic noise and proposed the methods to generate the sample-dependent and sample-independent semantic noise. Then, the framework of robust semantic communication system has been proposed to combat the semantic noise, where the adversarial training with weight perturbation has been developed. The masked VQ-VAE with noise-related masking strategy has been proposed as the architecture of the system and a discrete codebook for encoded feature representation has been designed. To improve the system robustness, the FIM that suppresses the noise-related and task-unrelated features has been proposed.  
Simulation results show that our proposed method can be applied in many downstream tasks and significantly improve the robustness of semantic communication systems against semantic noise with much reduced transmission overhead.

\bibliographystyle{IEEEtran}
\bibliography{IEEEabrv,Semantic}

\end{document}